\documentclass{article}

\usepackage{arxiv}

\usepackage[utf8]{inputenc} 
\usepackage[T1]{fontenc}    
\usepackage{hyperref}       
\usepackage{url}            
\usepackage{booktabs}       
\usepackage{amsfonts}       
\usepackage{nicefrac}       
\usepackage{microtype}      
\usepackage{lipsum}
\usepackage{hyperref}

\usepackage{todonotes}

\usepackage{fancyhdr}
\usepackage{graphicx,caption,subcaption}
\usepackage{amsmath}
\usepackage[toc,page]{appendix}
\usepackage{float}
\usepackage{multicol}

\usepackage{latexsym}

\usepackage{multicol,multirow}
\usepackage{amsmath,amssymb,amsfonts}
\usepackage{mathrsfs}
\usepackage{amsthm}
\usepackage{apacite}
\usepackage{rotating}
\usepackage{appendix}
\usepackage[authoryear]{natbib}
\usepackage{ifpdf}
\usepackage[T1]{fontenc}
\usepackage{times}
\usepackage{sourcesanspro}
\usepackage{newtxmath}
\usepackage{textcomp}%
\usepackage{xcolor}%
\usepackage{todonotes}
\newtheorem{alg}{Algorithm}

\title{MCMC for a hyperbolic Bayesian inverse problem in traffic flow modelling}

\author{
  Jeremie Coullon \\
  CEGE department\\
  University College London\\
  London, United Kingdon \\
  \texttt{jeremie.coullon@gmail.com} \\
  \and
    Yvo Pokern \\
  Department of Statistical Science\\
  University College London\\
  London, United Kingdon \\
  \texttt{y.pokern@ucl.ac.uk} \\
}

\begin{document}

\maketitle

\begin{abstract}
As a Bayesian approach to fitting motorway traffic flow models remains rare in the literature, we explore empirically the sampling challenges this approach offers which have to do with the strong correlations and multi-modality of the posterior distribution. In particular, we provide a unified statistical model to estimate using motorway data both boundary conditions and fundamental diagram parameters in LWR, a well known motorway traffic flow model. This allows us to provide a traffic flow density estimation method that is shown to be superior to two methods found in the traffic flow literature. To sample from this challenging posterior distribution we use a state-of-the-art gradient-free function space sampler augmented with parallel tempering.
\end{abstract}


\section{Introduction}
\label{section_introduction}


Fitting differential equations to data - a type of inverse problem - is an essential part of modelling in the sciences and engineering. It allows researchers to model complex systems to be able to understand and predict their behaviour. We can find applications in such varied fields as Geophysics \cite{Geophysics_PT}, Hydrogeology \cite{BIP_Darcy_flow}, and fluid mechanics \cite{BIP_fluids}.

In this paper we consider the Bayesain framework as outlined in \cite{Stuart_Acta_Numerica} for partial differential equations. We consider observations \(y\) generated by an observation operator \(\mathcal{G}\) polluted with noise \(\eta\):

\begin{equation}\label{general_BIP_equation}
y = \mathcal{G}(u) + \eta
\end{equation}

Starting from a prior belief on the parameter \(u\), the objective is to update this distribution based on observations represented by the likelihood.


For nonlinear PDEs such as the one studied in this paper, the posterior is unavailable analytically, and so one must resort to numerical methods such as sampling methods (Markov Chain Monte Carlo (MCMC), Sequential Monte Carlo (SMC)) (see \cite{Liu_MC_strategies} or \cite{Handbook_of_MCMC}). In this paper we use MCMC.


We consider as application vehicular traffic flow on motorways. Many approaches have been considered in the traffic flow literature to model such a system: systems of ODEs (\cite{Bando1995}, \cite{Gasser2004}), SDEs (\cite{stochastic_process_review}), or PDEs (\cite{Lighthill_whitham}, \cite{Zhang2}, \cite{A&R_resurection_of_fluid}). We focus on the most well known model, the Lighthill-Whitham-Richards (LWR) model, which is a conservation law with a nonlinear flux function:

\begin{equation}
\rho_t + (\rho V_e(\rho))_x = 0
 \end{equation}
 
With density \(\rho\) and \(V_e(\rho)\) the equilibrium density. We use subscripts to denote partial derivatives. Using the fundamental relationship relating flow \(q\) to density \(\rho\) and speed \(v\): \(q = \rho v \), we can define the Fundamental Diagram (FD) \(Q_e(\rho) = \rho V_e(\rho)\).
 
The main contributions of the paper are as follows: as a rigorous Bayesian treatment of motorway traffic flow models is rare in the literature, we explore empirically the sampling challenges these offer. In particular, we provide a unified statistical model to estimate both boundary conditions and fundamental diagram parameters in LWR. This allows us to provide a traffic flow density estimation method that is shown to be superior to two methods found in the traffic flow literature. Finally, we highlight how fitting these models is now tractable due to recent advances in gradient-free function space MCMC samplers.
 
The structure of the paper is as follows. In section \ref{section-TF_LWR} we give an overview of traffic flow theory, the LWR model, the traffic data, and previous work in this area. In section \ref{section-FD_only} we fit the Fundamental Diagram (FD) parameters directly to data using MCMC without considering the PDE model which is the usual approach in the traffic flow literature. In section \ref{sec:inverse_problem_methodology} we sample from the FD parameters and the boundary conditions within LWR. To overcome the sampling difficulties of this inverse problem we use a state-of-the-art gradient-free function space MCMC sampler and parallel tempering. We also compare this methodology to other methods both in the traffic engineering and statistical literature.

\section{Motorway traffic flow and the LWR model}
\label{section-TF_LWR}

\subsection{Traffic flow on the motorway}

\begin{figure}[ht]
  \centering
\includegraphics[width=.5\linewidth]{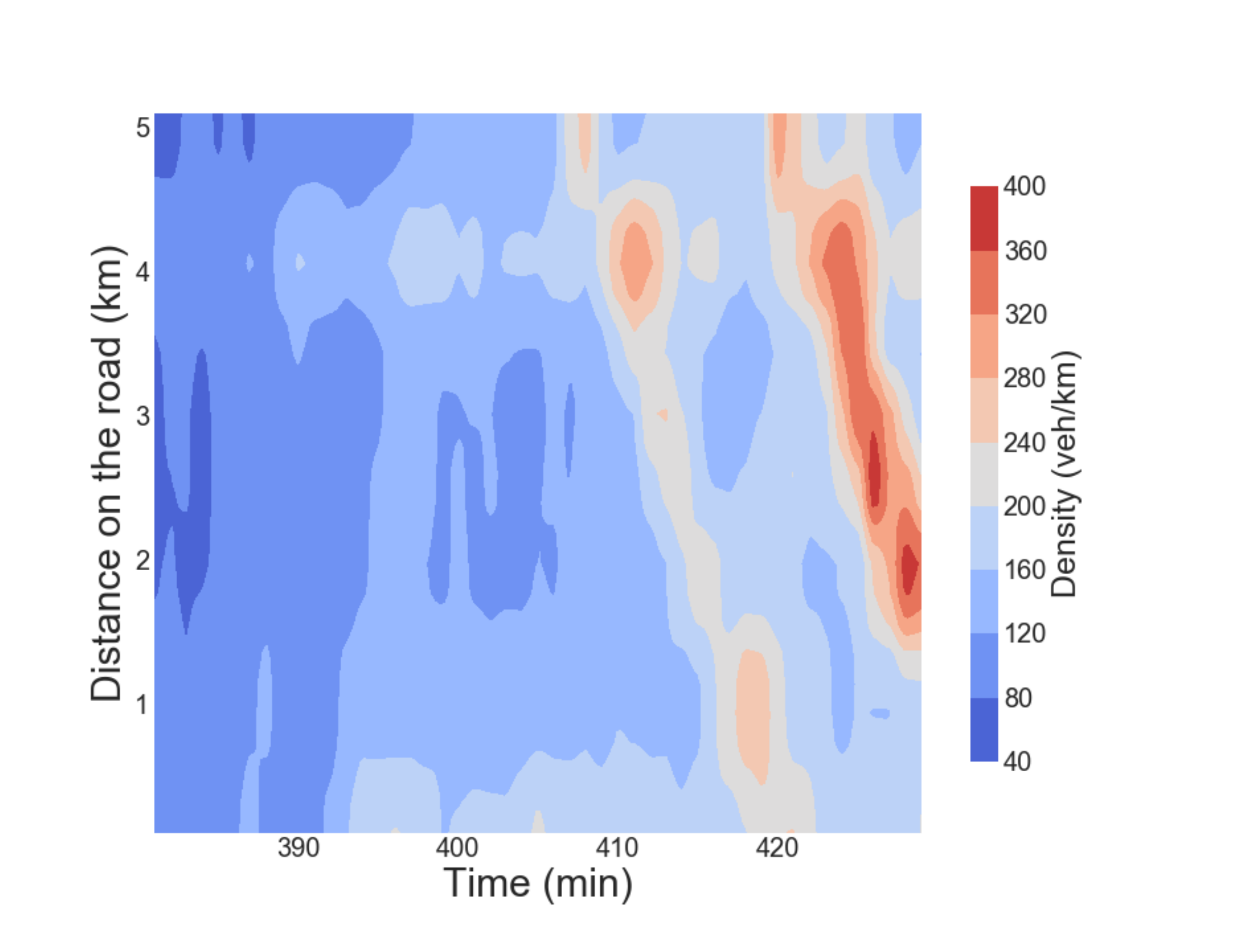}
  \caption{Density estimated from occupancy for the section of M25 on the 8th January 2007 between 6:21am and 7:09am. We observe forward moving free flow waves between minutes 381 and 405 which correspond to the movement of vehicles. We also observe backwards moving high density waves in the second half of the \(x-t\) plane.} 
  \label{intro_M25_data}
\end{figure}

We start by showing in figure \ref{intro_M25_data} traffic  density (the number of vehicles per km) in space and time for a 5 km stretch of M25 motorway in the UK for 49 minutes. In this plot time is denoted on the horizontal axis and space (distance on the road) on the vertical axis. Vehicles move from distance 0km to 5km (namely upwards in the plot) and forwards in time (namely to the right in the plot) so therefore move diagonally (upwards and to the right). We can see the vehicles' movement in the first few minutes (approximately between minutes 381 and 405) where they take a few minutes to cover the 5km section of road. This is consistent with a vehicle speed of approximately 120km/h (ie 2km/min). This vehicle movement corresponds to the free flow waves. We also observe backward moving waves in the second half of the \(x-t\) plot (from around minute 405 until the end) which are high density waves. These backwards moving waves correspond to the experience of needing to brake sharply when driving on motorways to avoid crashing into a traffic jam; congested flow waves move upstream in traffic.

\subsection{The Fundamental Diagram}

\begin{figure}[ht]
\centering
\includegraphics[width=.8\linewidth]{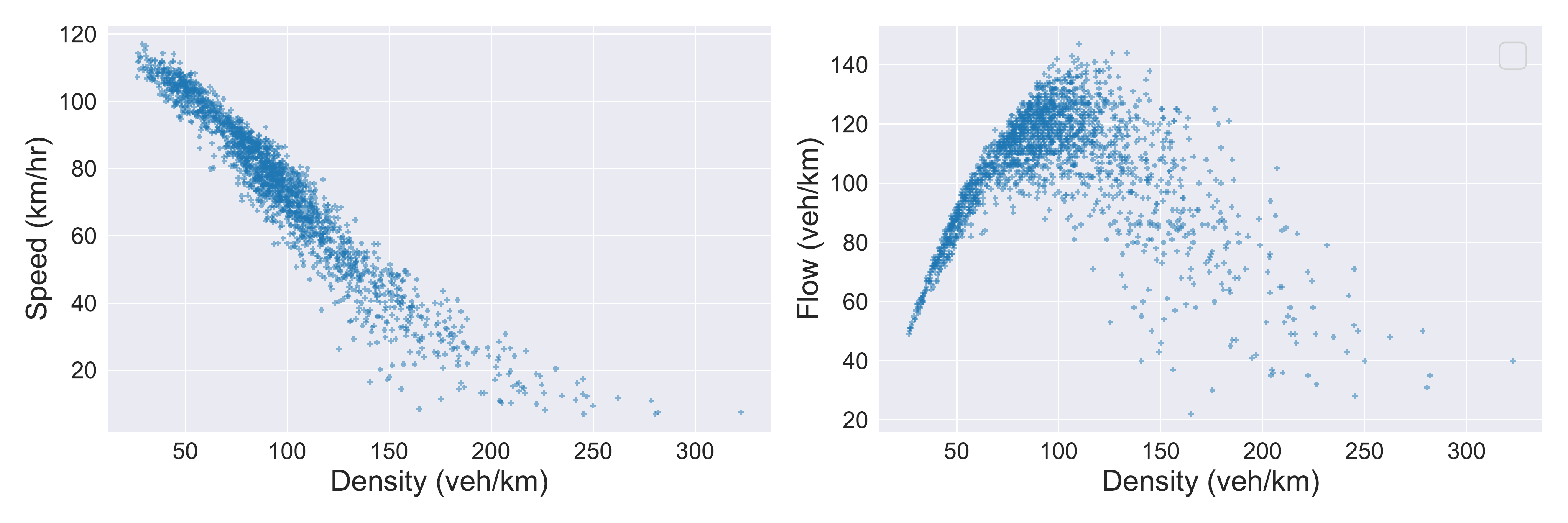}
\captionsetup{width=.7\linewidth}
\caption{Figures showing the empirical relationship between vehicle flow \(q\), density \(\rho\), and velocity \(v\). The data is taken from inductance loops over one minute averages from the M25}
\label{figure_fundamental_diagram}
\end{figure}

We now observe flow and density traffic data without the time component (flow being the number of vehicles per minute) as can be seen in figure \ref{figure_fundamental_diagram}, and noting that for the flow-density plot (the right-most plot) the measurements on the left of the plot can be well described by a curve with a positive slope. We define traffic following this curve to be in the state of \emph{free flow}. The rest of the points are scattered to the right of this curve which we call \emph{congested flow}. A main difference is that vehicles do not interact much in free flow; adding a vehicle does not decrease the speed of the other ones. In contrast, adding a vehicle to congested flow will decrease the speed of the other vehicles as they are forced to slow down to avoid collisions. We can see the effect of increasing density on average vehicle speed in the left-most plot of figure \ref{figure_fundamental_diagram}.

The fundamental diagram (FD), first studied in \cite{Greenshields_1934}  is the function used to described the relationship between density and flow as seen in figure \ref{figure_fundamental_diagram}. It is used to close the conservation of mass equation in PDE models of traffic, as will be described later.

Three variables are usually considered in the PDE approach of modelling traffic flow: density \(\rho\) which describes the number of vehicles per unit length, speed \(v\) which describes the average speed of vehicles at a point, and flow \(q\) which is the number of vehicles that pass a point in space during a unit of time. These three variables are linked through the following relationship:

\begin{equation}
q = \rho v
\end{equation}

\begin{figure}
\centering
\begin{subfigure}{.5\textwidth}
  \centering
\includegraphics[width=1\linewidth]{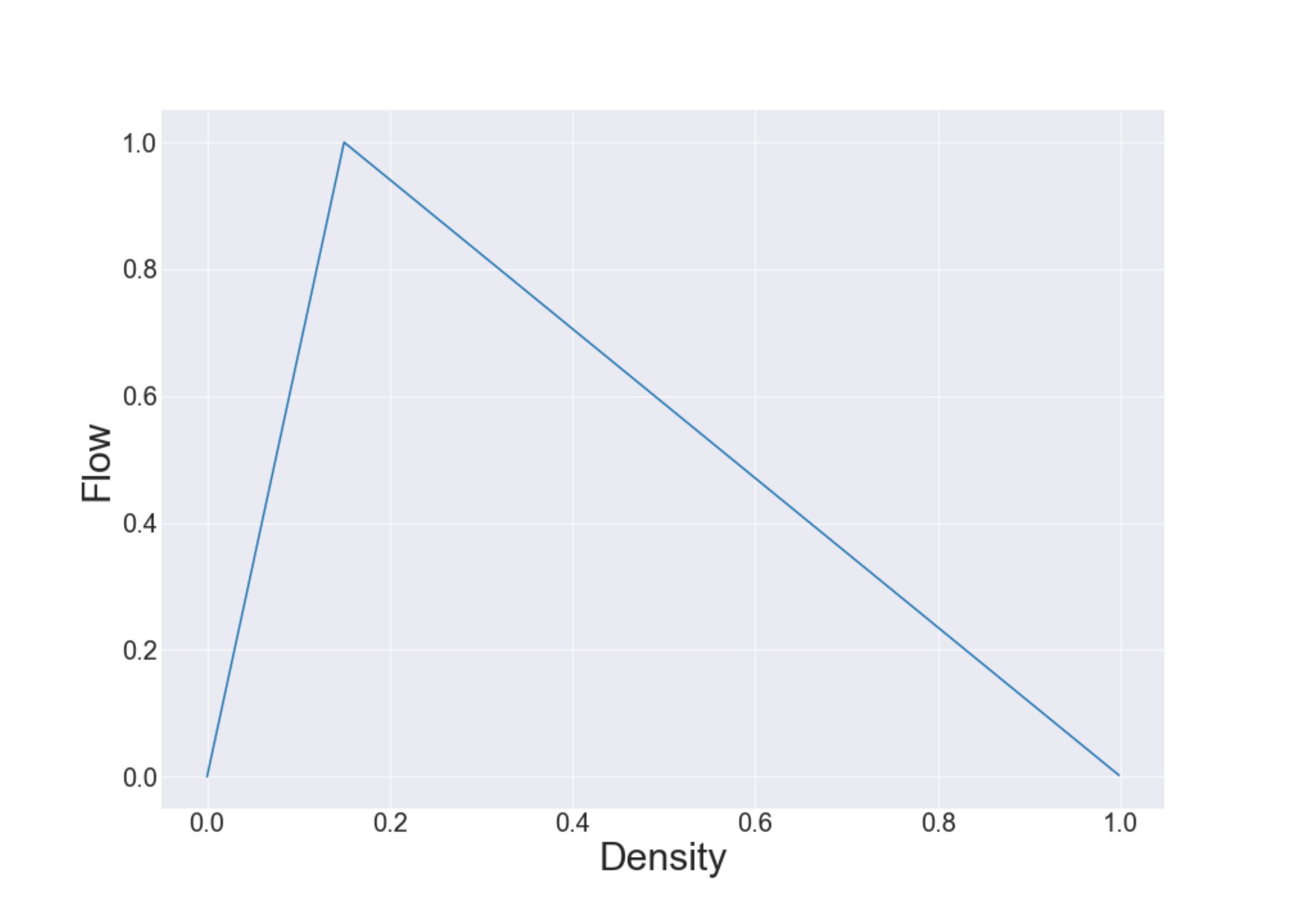}
  \caption{The Triangular FD}
  \label{FD_section-fig-triangular_FD}
\end{subfigure}%
\begin{subfigure}{.5\textwidth}
  \centering
\includegraphics[width=1\linewidth]{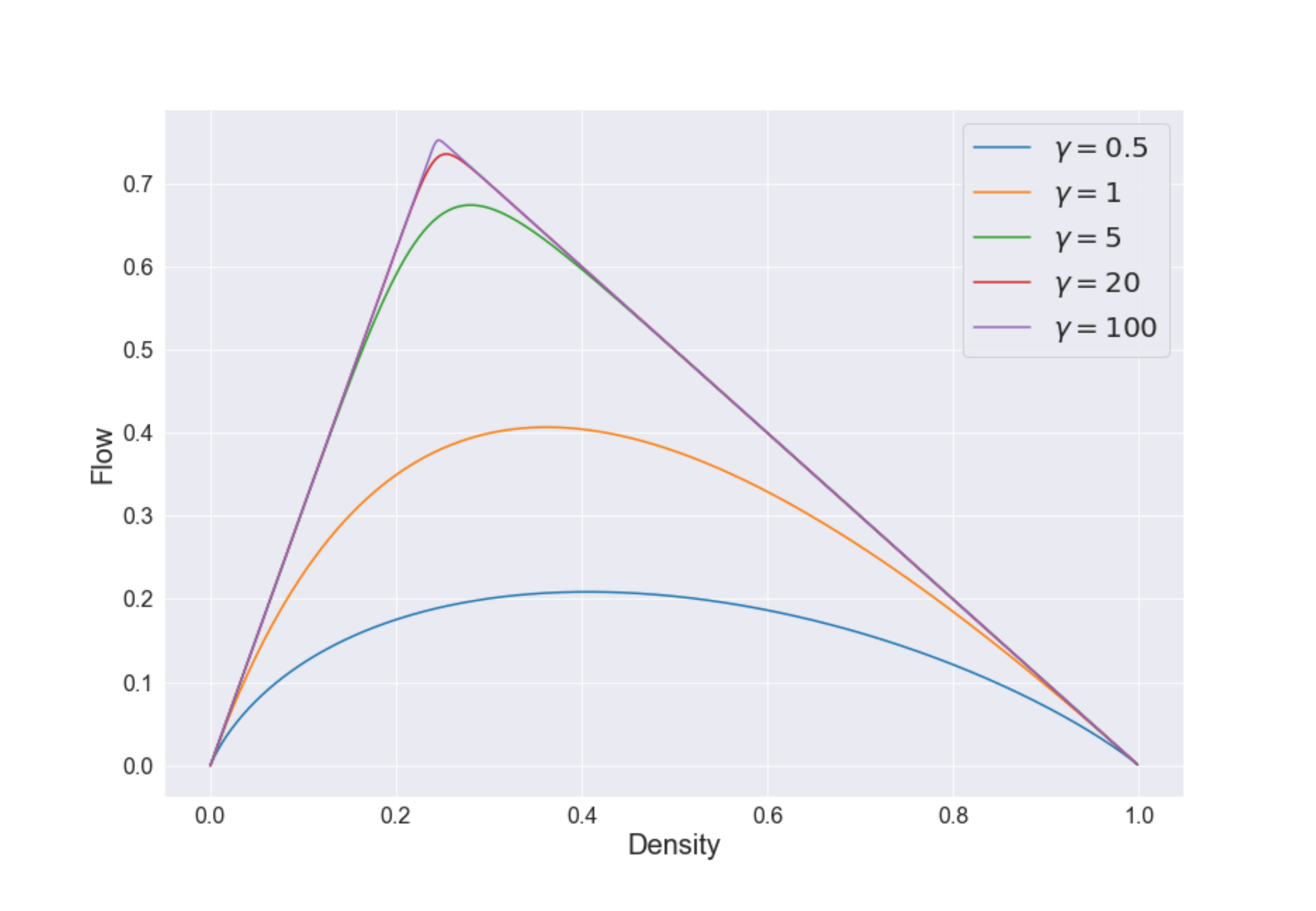}
  \caption{Del Castillo's FD}
  \label{FD_section-fig-delCastillo_FD}
\end{subfigure}
\captionsetup{width=.7\linewidth}
\caption{(a) Daganzo's triangular fundamental diagram plotted for dimensionless flow and density with \((q_c, \rho_c, \rho_j) = (1, 0.15, 1)\). (b) Del Castillo's fundamental diagram plotted for dimensionless flow and density with \((Z, \rho_j, u) = (1, 1, 3.1)\) and for \(\gamma \in [0.5, 1, 5, 20, 100]\)}
\label{FD_section-fig-triangular_vs_delCast}
\end{figure}

A popular example is Daganzo's \emph{Triangular FD} (also called the bi-linear FD) from \cite{Daganzo_CT_model} given in equation (\ref{triangular_FD_equation}) and plotted in figure \ref{FD_section-fig-triangular_FD}. This FD has important traffic flow quantities as parameters: the capacity \(q_c\) (maximum possible value of flow), the critical density \(\rho_c\) which separates free flow from congested flow, and the jam density \(\rho_j\) (maximum possible value of density). This FD is popular for its simplicity as well as for its computational efficiency when used in PDE models such as LWR (introduced in section (\ref{section_LWR_intro})). We will see later how the wave speed propagation of traffic flow models depends on the shape of the fundamental diagram.

\begin{equation}\label{triangular_FD_equation}
q(\rho) = \begin{cases}
\frac{q_c}{\rho_c}\rho \hspace{6mm} \rho < \rho_c \\ 
q_c \frac{\rho_j - \rho}{\rho_j - \rho_c} \hspace{6mm} \rho \geq \rho_c
\end{cases}
\end{equation}

We will focus in this paper on the Fundamental Diagram introduced by del Castillo in \cite{del_Castillo_FD_paper} called the negative power model (which we will simply call \emph{del Castillo's FD}). This FD has 4 parameters (all in \((0, \infty)\)): flow scaling term \(Z\), jam density \(\rho_j\), shape parameter \(\gamma\), and parameter \(u\) relating to the critical density. The shape parameter \(\gamma\) determines the tightness of the peak and in the limit of \(\gamma \to \infty\) we obtain the Triangular FD defined above. The parameter \(Z\) determines the vertical scaling of the FD, and \(u\) is related to the critical density via the following relation:

\begin{equation}
    \rho_c = \frac{1}{1 + u^{\gamma/(\gamma+1)}}
\end{equation}

We give the FD in equation (\ref{lit_review_DelCast_equation}) below and plot the FD for several values of \(\gamma\) in figure \ref{FD_section-fig-delCastillo_FD} to illustrate how del Castillo's FD includes the Triangular FD as a limiting case.

\begin{equation}\label{lit_review_DelCast_equation}
q(\rho) = Z \left[ (u\frac{\rho}{\rho_j})^{-\gamma} + (1 - \frac{\rho}{\rho_j})^{-w} \right]^{-\frac{1}{\gamma}}
\end{equation}

\subsubsection{LWR}
\label{section_LWR_intro}

As mentioned in section (\ref{section_introduction}), one of the ways to model traffic flow is to use partial differential equations (PDEs). This approach describes the macroscopic behaviour of traffic flow emerging from the individual interactions between vehicles, so we consider traffic flow as a continuum without representing the individual behaviour of vehicles. The differences between these models will therefore be in the assumptions made in how the system acts. We point to \cite{Bonzani_review} and \cite{Hoogendoorn2001} for a critical review on the topic.

An important class of these PDE models are those that consist of a single equation, which corresponds to a conservation law of the form:

\begin{equation} \label{conservation_of_vehicles}
\rho_t+(\rho v)_x=0
\end{equation}

As mentioned earlier, we use the subscripts \(\rho_t\) to denote \(\frac{\partial \rho}{\partial t}\) and \(\rho_x\) to denote \(\frac{\partial \rho}{\partial x}\). This conservation equation is derived from the obvious fact that vehicles are conserved on a length of road (assuming there are no on or off-ramps) (see \cite{FVMHP_book} for a derivation). Different assumptions in how the system acts will lead to different ways of closing this conservation equation. 

A standard way of doing this is to use the following form for the velocity: \(v = V_e(\rho)\) (see \cite{Coullon_phd} for an overview of PDE models in traffic flow). This assumes that vehicles adapts their speed instantaneously to a change in local density; i.e. traffic flow is always in equilibrium described by the function \(V_e(\rho)\). Multiplying this function by the density gives the fundamental diagram \(q=\rho V_e(\rho)\). This model is named the LWR model after Lighthill \& Whitham \citep{Lighthill_whitham} and Richards \citep{Richards} who developed it independently.

A property of the LWR model is that it allows the formation of \emph{shock waves} (discontinuities in the solution); more generally, nonlinear hyperbolic PDEs are prone to forming shocks. As a result, LWR with this choice of FD can model the propagation of upstream and downstream shock fronts on the highway. We can see these shock fronts in figure \ref{intro_M25_data}: the forwards moving waves occur for free flow traffic (namely for low density) and the backwards moving waves occur for congested flow traffic (namely for high density). 


In hyperbolic PDEs these shock fronts will move at speeds depending on the density and this is given by the gradient of the FD at that value of density. More generally, a disturbance in the initial condition of the PDE moves along \emph{characteristics}: curves in the \(x-t\) plane where the solution is constant (with speed also given by the gradient of the FD). So using the FDs in figure \ref{FD_section-fig-triangular_vs_delCast}, we can see that in free flow we have \(q'(\rho)>0\) so characteristics have positive speed, while in the congested regime they have negative speed. 

To solve LWR we need to choose the FD along with its parameters, and choose some inital conditions (density along the road at time \(0\)) and boundary conditions (density at the inlet and outlet of the road for the entire time period of the simulation).

\subsection{Numerical Method}
\label{section-numerical_method}

We would like to solve LWR for arbitrary FD parameters, initial conditions, and boundary conditions. However we cannot solve it analytically in the general case. We will use the open-source software Clawpack (\cite{Clawpack_software}) which is a package for solving conservation laws using finite volume methods (see  \cite{Clawpack_paper_citation} for information about the 5.0 release). Furthermore, an introduction to these finite volume methods along with an overview of the software can be found in \cite{FVMHP_book}. 

We give here an overview of the Godunov method (a first-order finite volume method) and point out a limitation. The idea of this method is to solve the PDE in conservation form (as in equation (\ref{Conservation_form_equation})), which ensures that the method behaves correctly in the presence of shock waves:

\begin{equation}\label{Conservation_form_equation}
\rho_t + f(\rho)_x = 0 
\end{equation}
We discretise space and time into cells, and consider methods of the form:

\begin{equation}\label{Godunov_equation}
\rho_i^{n+1} = \rho_i^n - \frac{\Delta t}{\Delta x}(F_{i+1/2}^n - F_{i-1/2}^n)
\end{equation}

With:
\begin{itemize}
\item \(\rho_i^n\) density at cell \(i\), time \(n\)
\item \(F_{i+1/2}^n\) flux (flow) at the the right boundary of cell \(i\) at time \(n\)
\item \(F_{i-1/2}^n\) flux (flow) at the the left boundary of cell \(i\) at time \(n\)
\item \(\Delta t\) and \(\Delta x\) the time and space discretisation
\end{itemize}

If we consider the density to be constant in each cell, we obtain a Riemann Problem at each boundary: an initial value problem with piecewise constant data and a single discontinuity. We then solve this Riemann Problem - which can be done analytically - at each boundary to find \(F_{i+1/2}^n\) and \(F_{i-1/2}^n\).

There are two main cases in the Riemann Problem:

\textbf{Case 1: \(\rho_{i-1}^n < \rho_i^n\)}

\begin{figure}[ht]
\centering
\includegraphics[width=.5\linewidth]{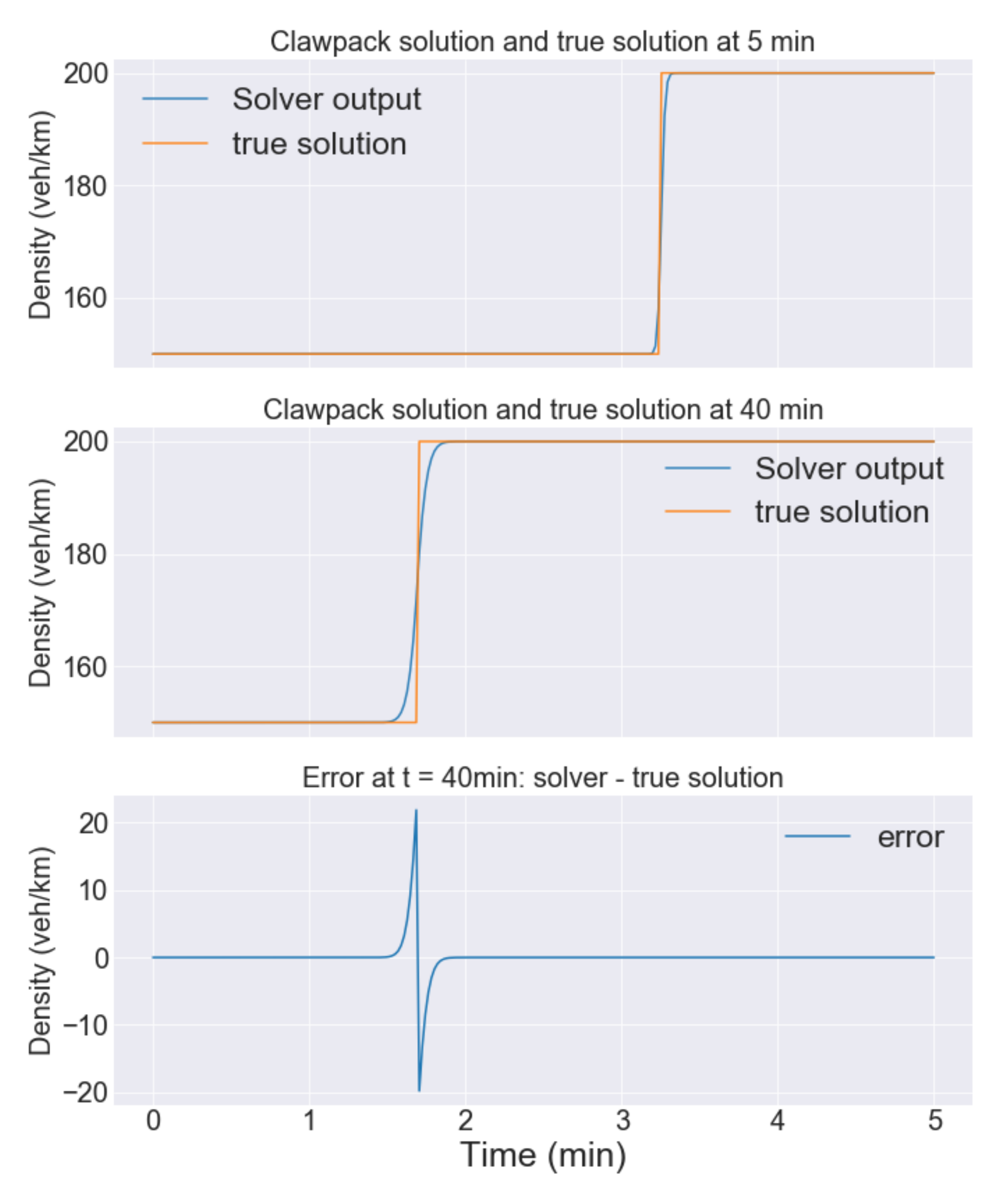}
\captionsetup{width=.7\linewidth}
\caption{We plot the analytic solution to the Riemann Problem along with its numerical solution using Clawpack. As time progresses we observe that the discontinuity is smoothed slightly. However, we notice that the position of the shock wave remains accurate.} 
\label{Clawpack_numerical_viscosity}
\end{figure}

In this case we will have a shock wave as seen in figure \ref{Clawpack_numerical_viscosity}: the discontinuity will simply be advected with speed: \(v_{shock} = \frac{f(\rho_i) - f(\rho_{i-1})}{\rho_i - \rho_{i-1}}\)  

\textbf{Case 2: \(\rho_{i-1}^n > \rho_i^n\)}

\begin{figure}[ht]
\centering
\includegraphics[width=.5\linewidth]{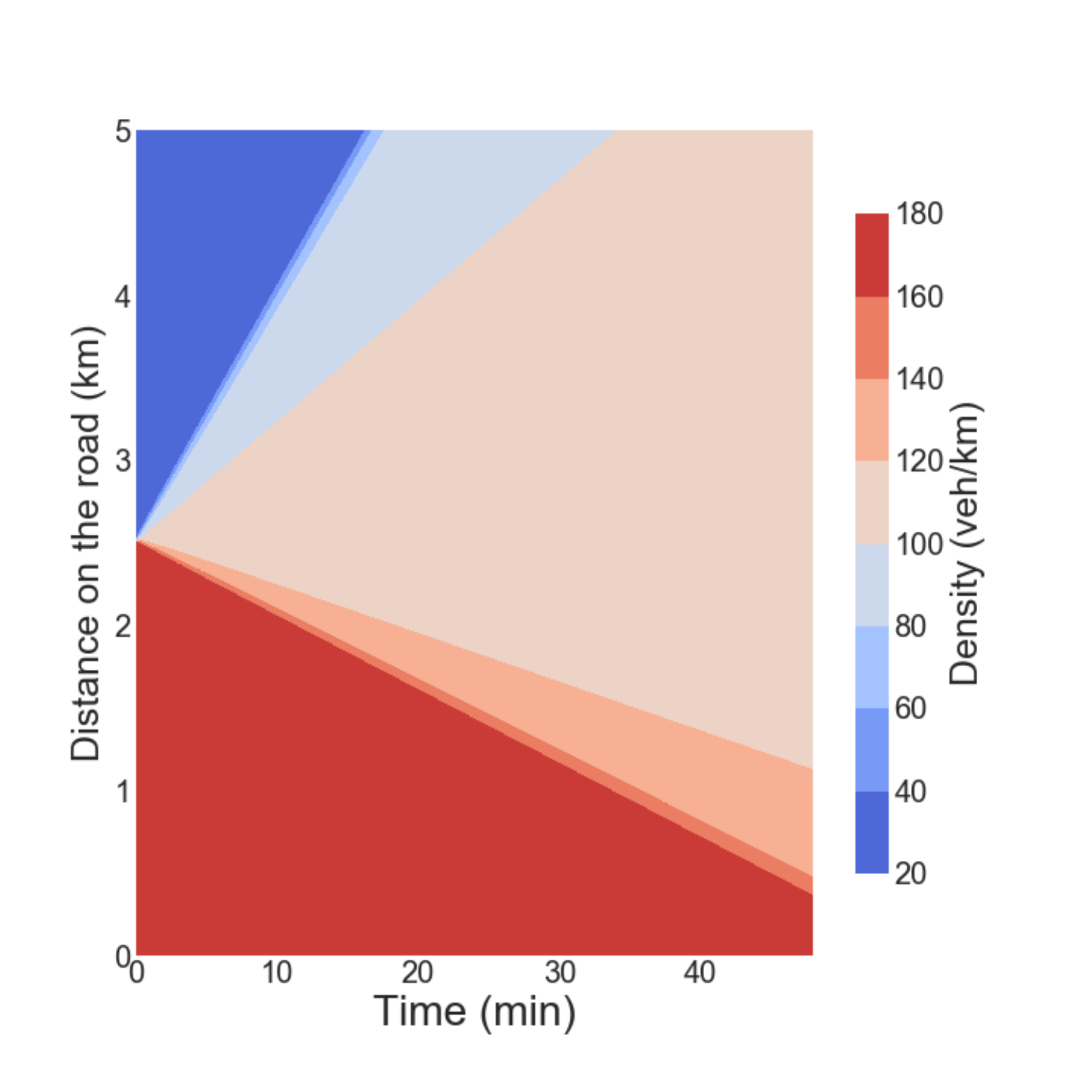}
\captionsetup{width=.7\linewidth}
\caption{The Riemann Problem with \(\rho_{i-1}>\rho_i\) causes a rarefaction wave. The initial condition (namely at \(t = 0\)) consists of a constant value of high density for \(x \in [0, 2.5]\) and a constant value of low density for \(x \in [2.5, 5]\). As the simulation moves forward in time we observe a rarefaction wave, or a fanning out of density values between the low and high values of the initial condition.} 
\label{Clawpack_rarefaction_xt}
\end{figure}

Here we will have a rarefaction wave as seen in figure \ref{Clawpack_rarefaction_xt}.

Creating methods that operate on the PDE in conservation form ensures that we can accurately model the position of the shock wave at any point in time and space. However they introduce a great deal of numerical viscosity that smooths out the solution. One can then extend Godunov's method to create second order numerical methods such as the Lax-Wendroff method (see \cite{FVMHP_book} for a detailed account) that models smooth solutions more accurately than the first order Godunov scheme but fails at discontinuities. To correct for this, one needs to add so-called \emph{flux limiters}. Using the Clawpack software, one only needs to solve the Riemann problem at each cell and the software automatically uses a Lax-Wendroff method with a flux limiter (the default one used is called the \emph{minmod} limiter).

However these methods still do exhibit some numerical viscosity. To illustrate this effect, we consider a Riemann problem (case 1 above) and compare the true solution of LWR with del Castillo's fundamental diagram for two different times to the output of Clawpack in figure \ref{Clawpack_numerical_viscosity}. This numerical viscosity is therefore expected to cause the posterior to be slightly different than if LWR was solved exactly. On the other hand, even though we observe jumps in density in motorway traffic, we do not expect them to be truly discontinuous: we rather expect them to be rapid but smooth transitions of density. Formally investigating the effect of this error would be an interesting area of research.


When defining the solver we must also choose the resolution \(\Delta x\)  and \(\Delta t\) such that the \emph{CFL condition} (Courant–Friedrichs–Lewy) - a necessary condition for convergence - is satisfied (see \cite{FVMHP_book}). To understand this condition, first we recall that as information in hyperbolic PDEs propagates with finite speed (along its characteristics) the solution \(\rho(x, t)\) is only affected by an interval of the initial condition. Points outside this interval do not affect the solution \(\rho(x, t)\). We define this interval to be the \emph{domain of dependence} of the solution \(\rho(x, t)\). The CFL condition then states that the numerical domain of dependence must contain the true domain of dependence (the numerical domain of dependence is similarly defined). For methods that only use adjacent cells to compute the solution at the next time step (for example equation (\ref{Godunov_equation})), this means that information must only come from the adjacent cells and not from cells further away. We then define \(\lambda_{max}\) to be the fastest wave speed of the PDE and we require \(\Delta t \leq \frac{\Delta x}{|\lambda_{max}|} \), ie:

\begin{equation}
\left| \frac{\Delta t \lambda_{max}}{\Delta x} \right| \leq 1
\end{equation}
This condition means we require the time resolution \(\Delta t\) to be smaller than the time it takes for the fastest wave speed to cross a cell of size \(\Delta x\). Thus information cannot propagate from non-adjacent cells to the current cell when calculating the density at the next time step.

The Clawpack software automatically chooses the time resolution for the solver based on the chosen spatial resolution and the CFL condition. We choose a spatial resolution of \(19\)m (as we have \(259\) cells for a \(5\)km length of road), and pass in a new boundary condition density value every \(1.5\)s (which corresponds to the time resolution required by CFL for typical FD parameter values found in data). Clawpack therefore chooses \(\Delta t \leq 1.5s\) depending on the FD parameter values. We used this resolution for the square wave test in figure \ref{Clawpack_numerical_viscosity}.

\subsection{Data}
\label{section_TF_LWR-data}

We use MIDAS data from the Highways Agency on the M25 in 2007. The data is measured on loop detectors spaced every 500m on the road which take measurements averaged every minute.  These loops measure count, occupancy, headway, and average speed. Count is the number of vehicles that have passed the detector in a minute, so therefore corresponds to flow (number of vehicles per unit time).  Occupancy (\(\omega_{occ}\)) is the percentage of time in a minute that the detector was recording the passing of a vehicle (so 100\% is gridlock, and 0\% means that no vehicles passed over the detector), and headway is the time difference between a vehicle leaving the detector and another one arriving. 

As density is a variable that must be included in models but is not directly measured by MIDAS detectors, we must estimate its value. One way to do this is to estimate it from speed data: for each lane, multiply the count by 60 and divide it by average speed (which is in km/h). We obtain the estimate \(\rho_{speed} := \frac{q}{v}\). The limitation of this approach is that it does not consider the size of vehicles (which can vary greatly; from small cars to lorries for example). 

Another approach is to estimate density from occupancy. To do this we use the relationship between occupancy and density (\cite{Heydecker_estimate_AVL}) \(\rho = \frac{\omega_{occ}}{ L}\), with \(L\) the average vehicle length. We use this to obtain density estimated from occupancy:

\begin{equation}\label{equation_density_occ}
    \rho_{occ} := \frac{\omega_{occ}}{L}
\end{equation}

There are several ways to estimate the average vehicle length \(L\) (or its reciprocal \(L^{-1}\) ): in \cite{Heydecker_estimate_AVL} the authors outline several methods to estimate \(L\) or \(L^{-1}\) and find that these have different trade-offs.
 
 However as we have the count data by vehicle type in the MIDAS data, we will use it to estimate the vehicle length at every minute. The flow by type is the overall count (for all lanes) of the number of vehicles that have passed the detector in the minute classified by type (type 1: 4m vehicles, type 2: 6m, type 3: 9m, type 4: 16m). As this count data is given over all lanes (rather than for each lane), we will have to assume that vehicles types are evenly distributed across lanes. Although this assumption allows us to estimate the average vehicle length in a practical way, it is unrealistic as motorways have lanes designated for slower vehicles (which includes longer vehicles). Using \(q_i\) to denote the count data for vehicles of type \(i\) and \(q\) the total count data we have: \( L = \left(4q_1 + 6q_2 + 9q_3 + 16q_4\right)/q\). We then use this value (calculated at every minute) to estimate density from occupancy for each lane: \(\rho_{occ} = \frac{\omega_{occ}}{L} \).

\begin{figure}[ht]
\centering
\includegraphics[width=.7\linewidth]{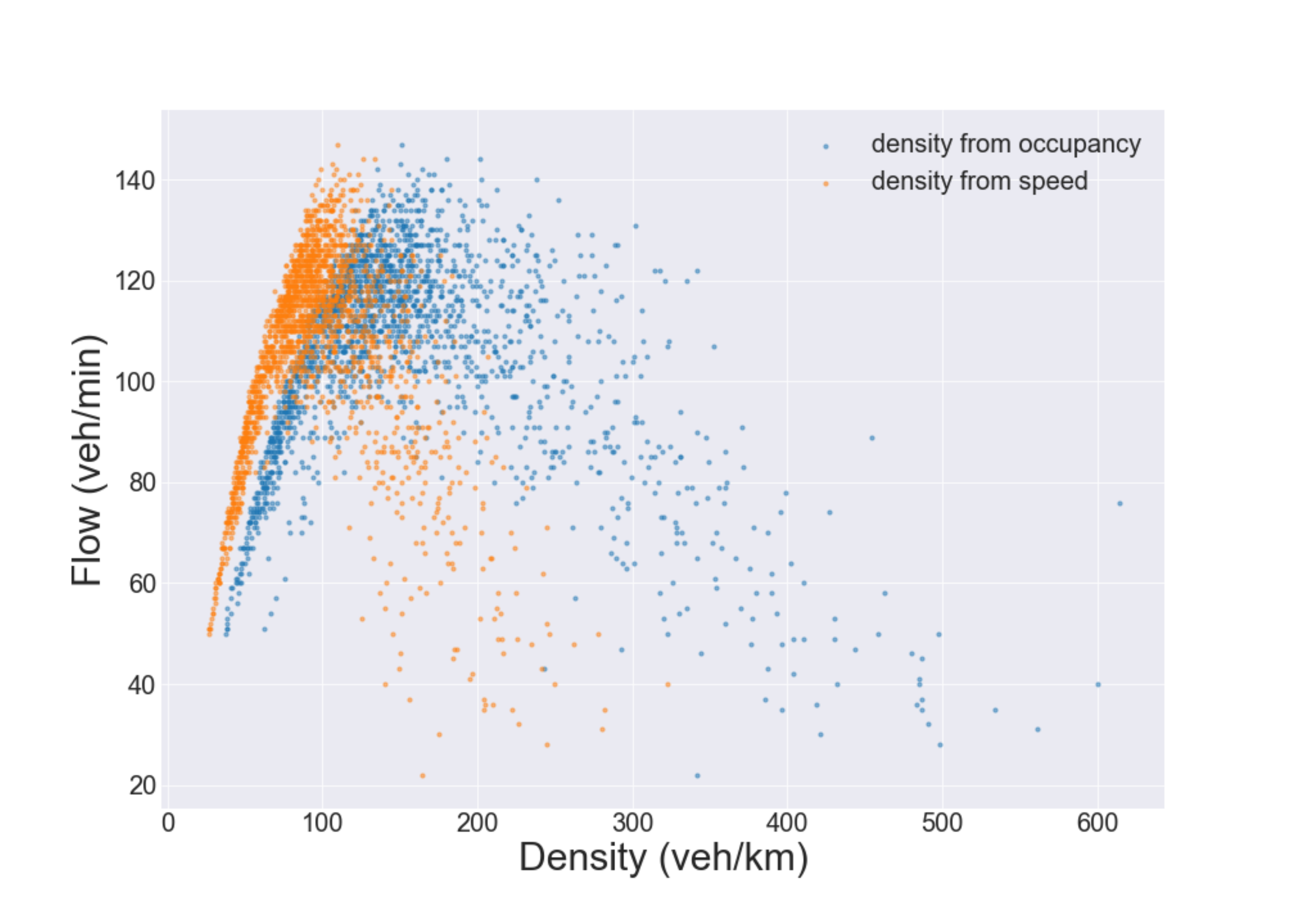}
\captionsetup{width=.7\linewidth}
\caption{Section of M25 on the 8th January 2007 between 6am and 10am. We plot flow vs density for two estimation methods: density from occupancy and density from speed summed over all lanes. These methods give very different estimates. In particular we note that the congested flow wave speeds vary greatly between methods, while the free flow wave speed is approximately the same.} 
\label{TF_den_occ_speed}
\end{figure}

We plot flow vs density using these two methods in figure \ref{TF_den_occ_speed} to show that they give very different results. In particular the congested flow wave speeds vary greatly between the two methods, while the free flow wave speeds are approximately the same. In section \ref{sec:results} we will estimate density in the BCs (as well as estimate the FD parameters) and compare the resulting wave speeds to those arising from density from speed and occupancy.

We will need to choose an appropriate section of road for analysis. As we are dealing with a single lane model, we need to choose a section of road with no in/out flows (junctions), the same number of lanes (which we will aggregate variables over),  few detector faults, and a consistent flow-density relationship. The chosen road is a 5km section of the M25, and as the detectors are spaced every 500m there are 11 detectors (the endpoints of the section are included). However some of the detectors have faults, so we use 8 detectors in our inference at the following locations (in km): \([0, 1, 2, 2.5, 3, 4, 4.5,  5 ]\). We use the detector measurements between 6:21am and 7:09am (including both endpoints) on the 8th January 2007; we therefore use 48min of data which corresponds to 49 time points.

\subsection{Previous related work}

A study close to the topic of this paper is a Bayesian analysis of traffic flow tested on motorway data by Polson and Sokolov in \cite{Polson_LWR}. The objective of the paper is to develop a methodology (using particle filtering) to estimate in real time traffic density and parameters in the LWR model. This allows real-time estimation of road capacity (maximum possible flow on the road) and critical density (density at which flow is maximised) that can adapt to the drop in capacity due to an accident on the road. We go over key points in their methodology and provide a critical review.

The motorway data they use includes occupancy \(\omega_{occ}\) which they use to estimate density (see formula \ref{equation_density_occ}). As explained in section \ref{section_TF_LWR-data}, they use this quantity to estimate density with the formula \(\rho = \frac{\omega_{occ}}{L}\) (see \cite{Heydecker_estimate_AVL} ) with \(\rho\) traffic density and \(L\) the average vehicle length. They use a constant average vehicle length, but it is unclear from the paper how it has been obtained.

They discretise the road into \(M\) cells and assume that the boundary conditions and initial conditions are known (they use density data from occupancy to construct these). Defining \(\theta_t = (\rho_{1t}, ... ,\rho_{Mt})\) to be the hidden state vector of traffic densities for each cell, the model they use in the particle filter is:

\begin{equation}\label{Polson_model}
\begin{cases}
y_{t+1} = H_{t+1}\theta_{t+1} + \epsilon_{t+1}^V \hspace{6mm} \text{with}\hspace{3mm} \epsilon_{t+1}^V \sim \mathcal{N}(0, V_{t+1})\\
\theta_{t+1} = f_{\phi}(\theta_t) + \epsilon_{t+1}^W \hspace{6mm} \text{with}\hspace{3mm} \epsilon_{t+1}^W \sim \mathcal{N}(0, W_{t+1})
\end{cases}
\end{equation}

Where \(V_t\) and \(W_t\) are evolution and equation error respectively, \(y_{t+1}\)  is the vector of measured traffic density, \(f_{\phi}\) is the LWR evolution equation with Fundamental Diagram parameters \(\phi\) calculated using a Godunov scheme (see \cite{FVMHP_book} for a comprehensive account of these numerical methods or recall section \ref{section-numerical_method}). The observation matrix \(H_{t+1}\) picks out the cells with the measurements. 
The objective of the methodology is to sample from the posteriors \(p(\theta_t | y^t)\) and \(p(\phi | y^t)\) with \(y^t = (y_1, ..., y_t)\) the current history of data. 

They consider a length of road on an interstate outside Chicago which seems to have two detectors separated by 845m. They discretise this distance into 4 cells (so each cell corresponds to a distance of 211m) and use a time discretisation of 5 minutes (which seems to be the resolution of data available). They run the analysis for 24 hours worth of data and estimate the drop in capacity due to a traffic accident. They also test their methodology on simulated data assuming known initial and boundary conditions. For this simulated data they use a spatial resolution of 300m and a time resolution of 5 seconds over a total road length of 1.5km and a time horizon of 1600s. 

As their objective is real time estimation of certain traffic flow quantities (such as traffic state and capacity) they use a particle filter rather than MCMC as it is more appropriate for real-time analysis. In contrast, we develop in this paper a more general methodology for estimating parameters in hyperbolic PDEs with a more rigorous treatment of the boundary and initial conditions. Indeed, they assume that the boundary and initial conditions are known using density estimated from occupancy. However estimating density from occupancy has problems (like all methods) which we summarise in section \ref{section_TF_LWR-data}; we will therefore impute the boundary conditions rather than estimate them directly from data.

Another issue with the methodology is the coarse discretisation of the LWR model (211m and 5min for space and time respectively for the analysis of real motorway data); at this resolution the numerical solver will rather coarsely approximate the underlying PDE. Moreover, the shock waves which are important features of these nonlinear PDEs will be slightly smoothed out, as we point out in section \ref{section-numerical_method}. Furthermore, each time step in the PDE solver includes a Gaussian error term as seen in equation \ref{Polson_model}; this could smooth out shock waves which could help the sampling methodology. However we point out that the numerical method which converges to the PDE (described in \cite{FVMHP_book}) does not include a random term. Adding a Gaussian error at each time step seems more like the discretisation of a stochastic PDE (the LRW model has no stochastic component), although it is unclear whether this discretisation would indeed converge to a specific SPDE under refinement of the grid. This may depend on the scaling of the variance in the Gaussian error with discretisation step. Moreover, adding a Gaussian error at each time step amounts to adding or removing a random fraction of a vehicle thus violating the conversation of total number of vehicles (ie: conservation of mass). The model used in the paper should therefore rather be considered as an ad-hoc discrete model inspired by the Godunov method for LWR rather than a discretisation of the PDE.

The methodology developed in this paper attempts to remedy these issues and be a more rigorous treatment of the Bayesian inverse problem. However, we reiterate that the objective of the paper described above is to develop a methodology for \emph{practical} real-time estimation of certain traffic quantities. Given this objective, these simplifications and assumptions may be justified.

\section{Parameter estimation in the FD: direct fit}
\label{section-FD_only}
In this section we consider fitting the FD directly to motorway data, namely without the PDE model. This corresponds to the usual approach in the engineering literature (see for example \cite{del_Castillo_FD_paper}).

\subsection{Statistical Model}

As the two methods to estimate density discussed in section \ref{section_TF_LWR-data} give very different results, we would like to build our likelihood based on a quantity that has fewer assumptions built in, namely flow (which is simply vehicle counts per minute). We use a Poisson model for the statistical error, which is a standard model for count data. We choose this model because it has the correct domain, is unimodal, and because of its simplicity. A drawback of this model is that inter-arrival times in a Poisson process are exponentially distributed, whereas we expect to not have a vehicle immediately following another one (especially for high speeds). However as the detectors count vehicles over a minute, many vehicles will have passed before the next count and the model misfit for small time resolutions should not be apparent.
Finally, the chosen section of road has four lanes, but our model is only for single-lane roads. We therefore sum flow values over all lanes to obtain the total number of vehicles, and if the individual flows are independent Poisson random variables, then the sum of flows also follows a Poisson distribution.

Letting \(\theta\) be the vector of FD parameters in LWR, and let \((\rho_i, q_i)\) (with \(i \in \{1, 2,..., N\}\)) be observed density and flow. We assume that data is iid with a Poisson error model which leads to the log-likelihood:

\begin{equation}\label{general_likelihood_equation}
l(\theta) \propto \sum_{i=1}^N (-\hat{q}_i(\theta) + q_i \log(\hat{q_i}(\theta)))
\end{equation}

With \(\hat{q}_i(\theta)\) the predicted flow for the \(i^{\text{th}}\) observation resulting from the observation operator \(\mathcal{G}(\theta)\) from equation (\ref{general_BIP_equation}). In this section we sample from the parameters in del Castillo's (equation (\ref{lit_review_DelCast_equation})), namely \((z, \rho_j, u, \gamma)\). Recall that the \(\gamma\) parameter controls the tightness of the peak, so as \(\gamma \to \infty\) the FD tends to the Triangular FD (equation (\ref{triangular_FD_equation}). In order to allow this parameter to take high values but while also allowing the Markov chain to converge, we invert this parameter and hence sample $\omega := \frac{1}{\gamma}$.

Based on domain knowledge of the realistic shape the FD can take, we set uniform priors for the FD parameters shown in equation (\ref{delCast_prior_values}) below:

\begin{equation}\label{delCast_prior_values}
\begin{cases}
z \sim \mathcal{U}(100, 400) \\
\rho_j \sim \mathcal{U}(300, 800) \\
u \sim \mathcal{U}(1, 10) \\
w \sim \mathcal{U}(0.004, 10) \\
\end{cases}
\end{equation}

Code for reproducing the sampling results in this paper can be found at \url{https://github.com/jeremiecoullon/BIP_LWR-paper}.

\subsection{Direct Fit}

Here we fit del Castillo's FD directly to flow-density data. So the predicted flow \(\hat{q}_i(\theta)\) in equation (\ref{general_likelihood_equation}) is simply flow predicted from observed density \(\rho_i\) using del Castilo's FD with parameters \(\theta = (z, \rho_j, u, \omega)\) (so without using the LWR model). This is a usual approach in the traffic flow literature (see for example \cite{del_Castillo_FD_paper}).

\begin{figure}
\centering
\includegraphics[width=.6\linewidth]{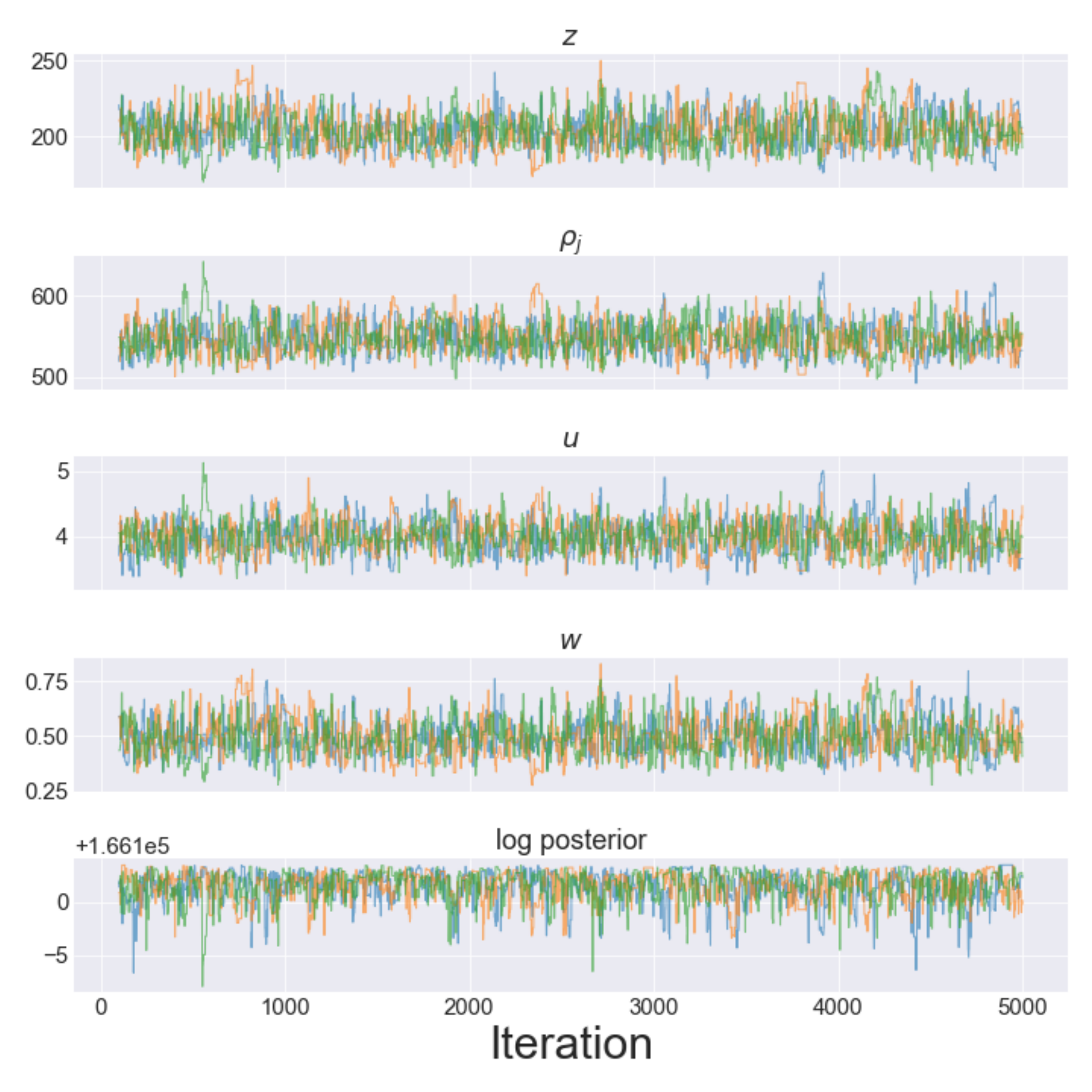}
\captionsetup{width=.7\linewidth}
\caption{Posterior samples from a direct fit of del Castillo's FD to M25 data. Trace plots of sampled parameters against flow-density data. The 3 colours correspond to the 3 MCMC chains }
\label{FD-direct_fit_trace}
\end{figure}

We sample from the posterior using a random walk Metropolis Hastings algorithm. We run 3 chains for 5K iterations with covariance matrix given in the appendix in section (\ref{appendix-FD_sampling}). We obtain acceptance rates of \(24.4\%\), \(23.1\%\), and \(23.8\%\), and show the trace plots for the parameters in figure \ref{FD-direct_fit_trace} which suggest good mixing. 

\begin{figure}
\centering
\begin{subfigure}{.5\textwidth}
  \centering
\includegraphics[width=1\linewidth]{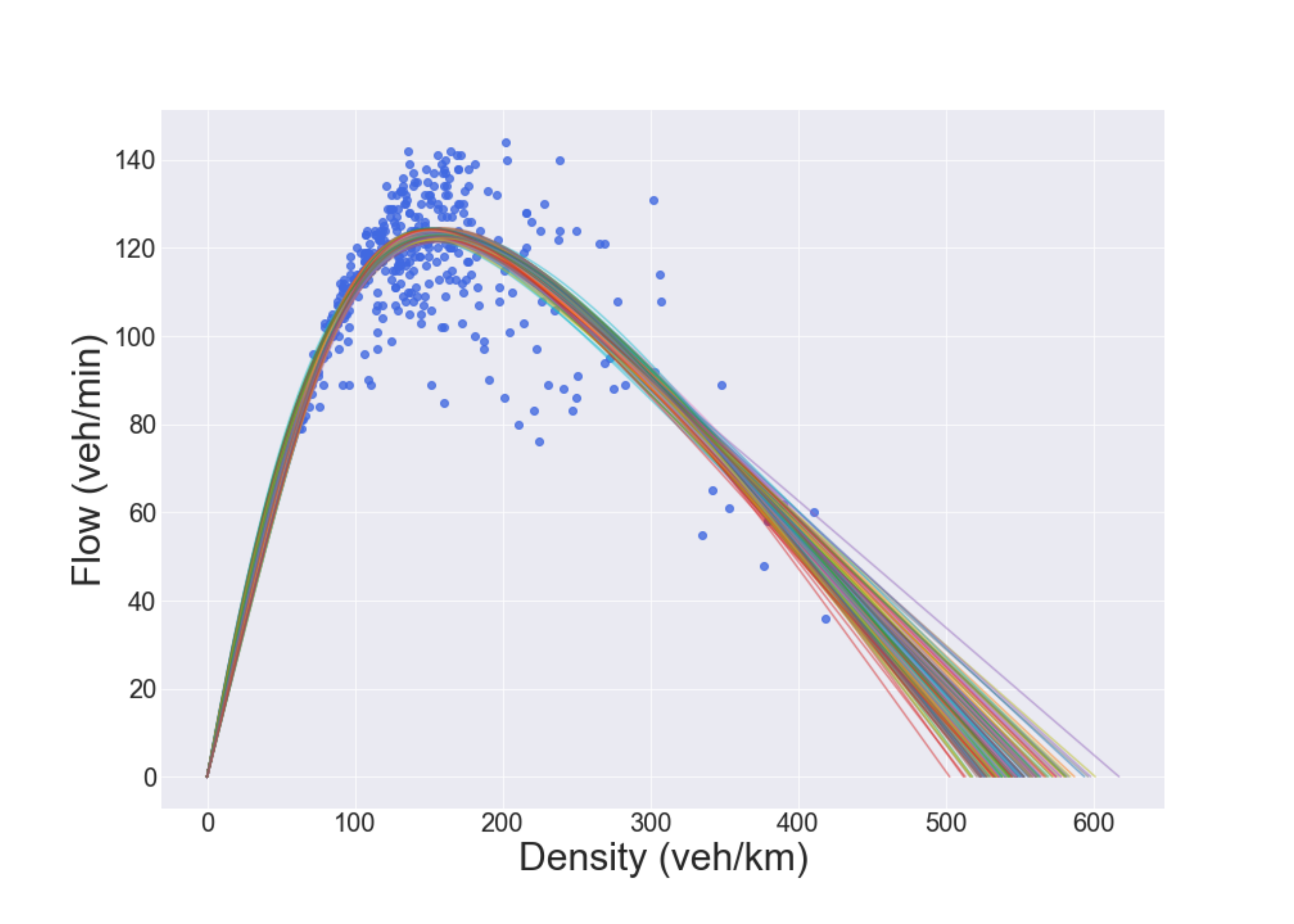}
  \caption{FD samples}
  \label{FD-direct_fit_samples}
\end{subfigure}%
\begin{subfigure}{.5\textwidth}
  \centering
\includegraphics[width=1\linewidth]{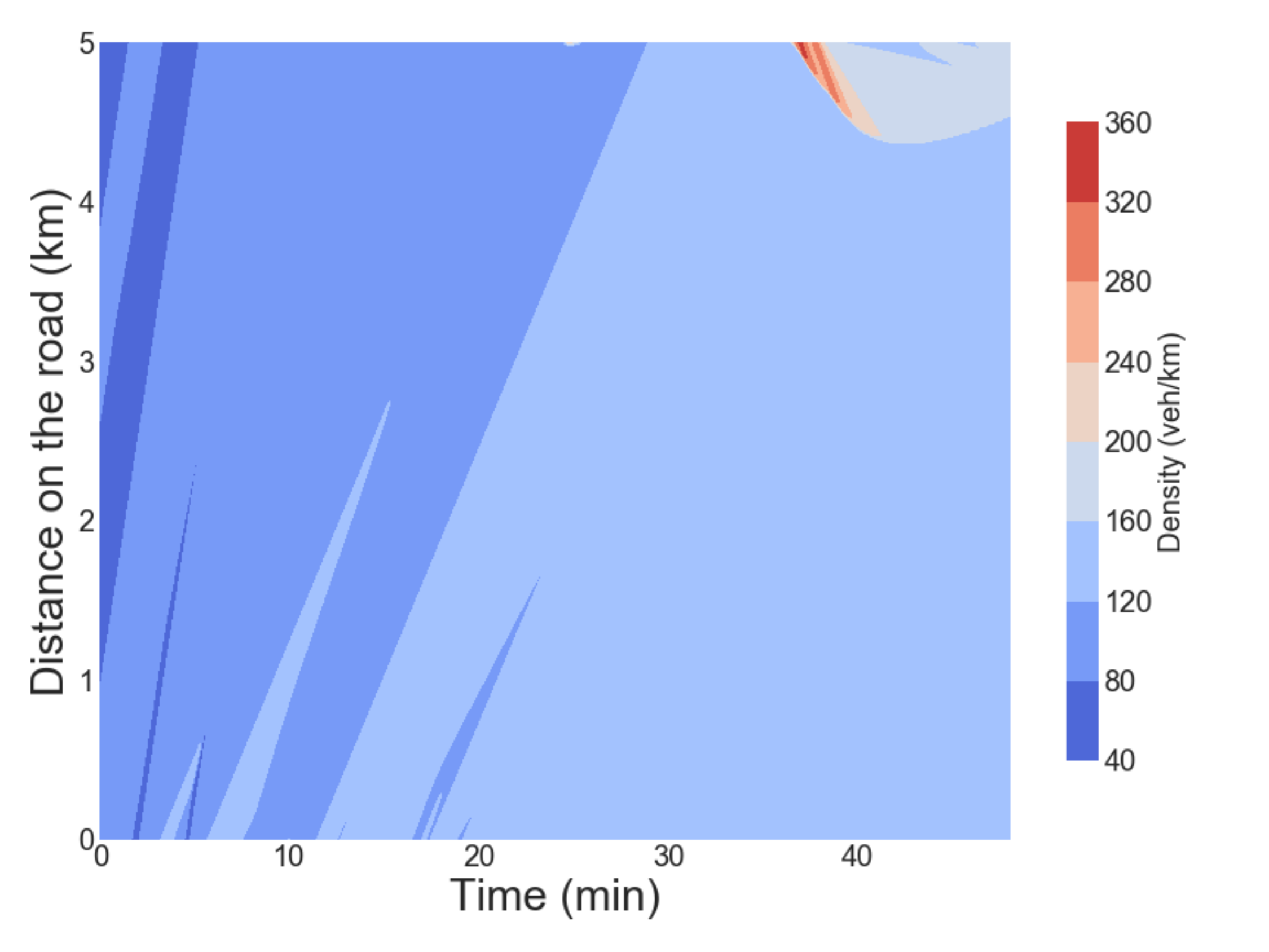}
  \caption{Solution in the $x-t$ plane from the posterior mean parameters}
  \label{FD-direct_fit_XT}
\end{subfigure}
\captionsetup{width=.7\linewidth}
\caption{Posterior samples from a direct fit of del Castillo's FD to M25 data. (b) Plotted FDs using the samples (c) Density in the \(x-t\) plane from LWR. Parameters used are the posterior mean from samples. We notice that the congested flow waves do not cross the domain as they do in the data}
\label{FD-direct_fit}
\end{figure}

We show FD plots with sampled parameters against flow-density data (using density from occupancy) in figure \ref{FD-direct_fit_samples}. We then show the output from LWR in the \(x-t\) plane using the posterior mean samples in figure \ref{FD-direct_fit_XT}. Comparing this plot to the real data in figure \ref{intro_M25_data} we can clearly see that the congested flow wave speed is incorrect (namely, the congested flow waves in the case of the direct fit do not cross the domain at all). This suggests a misfit of the model.

\section{Inverse problem methodology}
\label{sec:inverse_problem_methodology}

In this section we infer the parameters in the LWR model to predict flow values $q_i$ in equation \ref{general_likelihood_equation}.  The observation operator $\mathcal{G}$ solves the LWR model given some boundary conditions, initial conditions, and the FD parameters $\theta$, maps the density output to flow using the FD with parameters $\theta$, and picks out the flow at the $(x,t)$ values corresponding to observations.

Estimating the parameters in this way capture the dynamic behaviour of the PDE compared to the direct fit, Furthermore, estimating the BCs accounts for the uncertainties in the density estimates which were discussed in section \ref{section_TF_LWR-data}. We therefore aim to estimate the 4 FD parameters along with the 2 BCs.

The BCs are functional parameters so function space samplers are required for this. This is an active area of research which has yielded many samplers and methods such as random walk \citep{Cotter_pCN} $\infty$-MALA, $\infty$-HMC \citep{Beskos_geometric_MCMC_2017}, and Likelihood Informed Subspaces \citep{Cui_LIS}. However most efficient samplers require gradients which are not available in our application; we must therefore restrict ourselves to gradient-free samplers. The first gradient-free sampler to try is pCN \citep{Cotter_pCN} which is the function space extension of random walk. This algorithm is simple to implement but will mix slowly if the parameters are correlated or if the posterior is multi-modal. This is the case in our application (see section \ref{sec:multimodality} for a discussion): the FD and BC parameters are highly correlated and the BCs exhibit multi-modality. We therefore use the functional ensemble sampler (FES) \citep{Coullon_FES} which is a gradient-free sampler that can handle correlations in the parameters. We extend it with a parallel tempering scheme to mix efficiently within the modes.

In the following sections we describe the sampler along with the BC prior and the treatment of the initial conditions in LWR.

\subsection{Functional Ensemble Sampler}

The functional ensemble sampler (FES) \citep{Coullon_FES} combines pCN and the affine invariant ensemble sampler (AIES) \citep{Goodman_2010}. For exposition purposes we describe the sampler for a posterior distribution that only includes a single functional parameter. The extension to our application (two functional parameters and a finite dimension parameter) is straightforward.

To apply this sampler, we first calculate the Karhunen–Lo{\`e}ve (KL) expansion for the Bayesian prior distribution, assumed to be a Gaussian process prior. We then use a Metropolis-within-Gibbs sampler that uses AIES to sample the posterior distribution on the low-wavenumber KL components and uses pCN to sample the posterior distribution of the high-wavenumber KL components. Alternating between AIES and pCN updates, we obtain an efficient functional sampler without requiring detailed knowledge of the target distribution.

We give the pseudocode for the algorithm below:

\begin{alg}[Functional ensemble sampler]{\label{alg:fes_algorithm}}
~
\newline
To sample a distribution $\pi(\mathop{du}) \propto \exp\left(\phi(u)\right) \pi_0(\mathop{du})$
where
$\pi_0 = \mathcal{N}\left(0, C\right)$, perform the following steps:
\begin{enumerate}
\item Identify a matrix $J$ whose columns are the first $M$ eigenvectors of $C$.
    Set $P = J J^T$
    and $Q = I - J J^T$.
\item Initialize an ensemble of walkers $\left(X_1^0, ... X_L^0\right)$.
\item For $\tau = 0, 1, \ldots$:
\begin{enumerate}
    \item For $i = 1, \ldots, L$: 
    \begin{enumerate}
    \item Randomly choose a walker $X_j^{2\tau} \neq X_i^{2\tau}$.
    \item Propose the update         \begin{equation}\label{eq:FES_alg_AIES_proposal}
            \tilde{X}_i^{2\tau} = X_i^{2\tau} + \left(1 - Z\right) P \left(X_j^{2\tau} - X_i^{2\tau}\right),
    \end{equation}
    where $Z \in \left[1\slash a, a\right]$ has density $g\left(z\right) \propto 1 \slash \sqrt{z}$.
    \item
    Set $X_i^{2\tau} = \tilde{X}_i^{2\tau}$ with probability 
    \begin{equation}
        \min\left\{1, Z^{M-1} \frac{\pi\left(\tilde{X}_i^{2\tau}\right)}{\pi\left(X_i^{2\tau}\right)}\right\}.
    \end{equation}
    \end{enumerate}
    \item Set
    $\left(X_0^{2\tau + 1}, \ldots, X_L^{2\tau + 1}\right) = \left(X_0^{2\tau}, \ldots, X_L^{2\tau}\right)$.
    \item For $i = 1, \ldots, L$:
    \begin{enumerate}
    \item Propose the update
    \begin{equation}\label{eq:FES_alg_PCN_proposal}
        \tilde{X}_i^{2\tau + 1} = P X_i^{2\tau + 1} + Q \left(\sqrt{1 - \omega^2} X_i^{2\tau + 1} + \omega \xi\right),
    \end{equation}
    where $\xi \sim \mathcal{N}\left(0, C\right)$.
    \item Set $X_i^{2\tau+1} = \tilde{X}_i^{2\tau+1}$ with probability
    \begin{equation}
        \min\left\{1, \exp\left(\phi\left(\tilde{X}_i\right) - \phi\left(X_i\right)\right)\right\}.
    \end{equation}
    \end{enumerate}
    \item Set
    $\left(X_0^{2\tau + 2}, \ldots, X_L^{2\tau + 2}\right) = \left(X_0^{2\tau + 1}, \ldots, X_L^{2\tau + 1}\right)$.
\end{enumerate}
\end{enumerate}
\end{alg}

The main tuning parameter  in  FES is $M$,  which controls how many KL coordinates are included in the AIES sampling. \cite{Coullon_FES} recommend keeping this parameter less than $20$ as the performance of AIES tends to deteriorate for high dimensional distributions. The other parameter to tune is the pCN step size $\omega$. Finally $a$ is also a tunable parameter but as recommended in \cite{Coullon_FES} we fix it to $a=2$.

\subsection{Parallel Tempering}

\emph{Parallel Tempering} (PT), also known as Replica Exchange MCMC (\cite{Liu_MC_strategies}, \cite{Handbook_of_MCMC}) is an algorithm used to sample from multimodal distributions. We augment the state space by introducing an additional discrete parameter \(\beta_{temp} \in  I _{temp} =\{ \beta_1=1 < \beta_2 <... < \beta_L\}\) with \(\beta_i \in [0,1]\). Defining the unnormalised target posterior as \(\pi(x) \propto \exp\{ - k(x) \}\) we can augment the state space by tempering the likelihood (with \(-\phi(x)\) the log-likelihood, \(-\phi_0(x)\) the log-prior, and \(c(\beta_{temp})\) the pseudo-prior (as defined in \cite{Handbook_of_MCMC})):

\begin{equation}
\pi(x, \beta_{temp}) \propto \exp \{ -\beta_{temp}\phi(x) - \phi_0(x) \} c(\beta_{temp})
\end{equation}


We therefore run \(L\) chains in parallel to target the joint distribution \(\Pi(x_1, ... , x_L) = \pi_1(x_1), ... , \pi_L(x_L)\)  where \(\pi_i(x)\) is the posterior tempered using the inverse temperature \(\beta_i\).

We define \(\alpha_0\) to be the probability of making a within-temperature move, sample \(U \sim \mathcal{U}(0,1)\), and run the algorithm:

\begin{itemize}
\item if \(\alpha_0 > U\), perform a within-temperature move; this can be done using any MCMC sampling scheme
\item if \(\alpha_0 \leq U\), choose uniformly a pair of posteriors \(\pi_i(x_i)\) and \(\pi_j(x_j)\) (usually chosen so that the inverse temperatures are adjacent) and swap the states \(x_i\) and \(x_j\) with probability \(\alpha = \min\{ 1, \frac{\pi_i(x_j) \pi_j(x_i)}{\pi_i(x_i) \pi_j(x_j)} \}\).
\end{itemize}

Many extensions of these tempering algorithms have been proposed in the literature, such as ones generalising the between-temperature moves in \cite{Quant_TA_paper} or ones defining a continuous temperature schedule in \cite{continuously_tempered_HMC}.

A benefit of this algorithm is that the chains can be run on parallel cores rather than sequentially. However, one needs to choose a temperature schedule \(I_{temp}\). To tune this schedule we use the following iterative tuning procedure (based on the one outlined in \cite{Atchade_optimal_scaling}): we raise the likelihood to the power \(\beta_{temp}\) with \(\beta_{temp} \in [0,1]\), and find a value of \(\beta_{temp}\) that allows the chain to mix well. We then find a colder temperature such that the swap acceptance rate is approximately \(23\%\). We then fit a geometric schedule between these two values and extrapolate to find the other temperatures. We then check that the acceptance rate for temperature swaps is around \(23\%\) for these spacings.


\subsection{Functional ensemble sampler with parallel tempering}

The sampler used in our application merges FES and PT and includes some modifications on FES as shown in algorithm \ref{alg:fes_algorithm}. The details of our implementation are as follows:

\begin{itemize}
    \item We set $M=4$ and so the low dimensional space is $12$ dimensional: $4$ for each BC and $4$ for the FD.
    \item The Metropolis-within-Gibbs sampler alternates 4 steps rather than 2: the AIES update for the BCs and FD, the pCN update for the inlet BC, the pCN update for the outlet BC, and the temperature swaps.
    \item We use a random scan Metropolis-within-Gibbs sampler rather than deterministic scan, and we tune the move probabilities to obtain good mixing. The tuned parameters can be found in the appendix.
    \item We use 13 walkers and use the parallelised version of FES (see \cite{Coullon_FES}).
    \item We use a uniform prior for the FD parameters and a "log-OU" prior for the BCs as described in section \ref{section-BC-prior_elicitation} below.
\end{itemize}

\subsection{BC Prior elicitation}
\label{section-BC-prior_elicitation}


In this section we describe how we chose a prior for the two boundary conditions (BCs). As density from occupancy (see section (\ref{section_TF_LWR-data})) is considered an appropriate method for estimating density, we will use it to elicit the prior. As discussed previously, it is estimated using the average vehicle length (unlike density estimated from speed). Eliciting the prior in this way will encode our prior belief that the estimated BCs should not deviate too far from density estimated from occupancy.

We will use time series of density from our section of road to fit a Gaussian process prior for the BCs (of course discarding the day that we use in inference).

We choose as prior a "log-OU" process. By that we mean that the logarithm of the centered BCs follow an Ornstein Uhlenbeck (OU) process. We choose this prior for 3 reasons:

\begin{itemize}
    \item We would like density to always be positive.
    \item We would like the prior to allow sudden excursions in density corresponding to high density waves. Indeed, with a log-OU prior we model the logarithm of the centered BCs with an OU process. As a result, a high value of density will a priori have a higher variance which enables high density waves, and a low value of density will a priori have a low variance. 
    \item We would like to be able to sample easily from the prior as well as evaluate the probability of a sample under the prior. This is computationally inexpensive to do with a log-OU prior.
\end{itemize}

We first give a succinct overview of the OU process and then estimate the parameters of this process from traffic flow data.

For a given BC (ie: inlet or outlet), let \(Y_t\) be the logarithm of the BC at time \(t\) and let \(X_t := Y_t - \mu(t)\) with \(\mu(t)\) be the time-dependent mean. Then we choose \(X_t\) to be the unique solution of the stochastic differential equation \(dX_t = -\beta X_t dt + \sigma dW_t\) (with \(W_t\) a Wiener process), with \(\beta>0\) and \(\sigma>0\) the mean-reversion parameter and diffusivity parameter respectively (see \cite{simulation_inference_SDE}).

We fit an OU process to centred log-BCs for the inlet and outlet BC together, as fitting them seperately yielded similar OU parameters.  The inlet BC is a function of time that returns density, and corresponds to the inlet of the studied stretch of road (ie: \(x=0\)) . Similarly, the outlet BC corresponds to the outlet of the road (ie: \(x=5\)km).

For the inlet and outlet detector data we keep only weekdays, discard the 1st January, and keep only the 75 and 65 first days for the inlet and outlet detectors respectively. We removed these last days as they have unusual density curves. We also removed the 8th January as this is the dataset used in the inference. We fit the mean \(\mu(t)\) to the logarithm of traffic curves \(Y_t\), and then fit the OU parameters \(\beta\) and \(\sigma\) to \(X_t\) (we fix \(\Delta t=1\) to define a unit of time to be \(1\) minute). We apply a very slight smoothing to the log-BC means to ensure that they are smooth.

We estimate the parameters for the inlet and outlet together using MCMC. Defining \(X^i\) to be the i-th measured density curve and \(\Lambda\) the precision matrix for the OU process, we write the likelihood as:

\begin{equation}
l(\Lambda)  \propto -\frac{N}{2}\log|\Lambda| - \frac{1}{2}\sum^NX^i\Lambda X^i
\end{equation}

We use a flat prior for the parameters and use a random walk Metropolis sampler to sample from the posterior.  The posterior mean is \(\beta=0.22\) and \(\sigma=0.256\).

\begin{figure}
\centering
\begin{subfigure}{.5\textwidth}
  \centering
\includegraphics[width=1\linewidth]{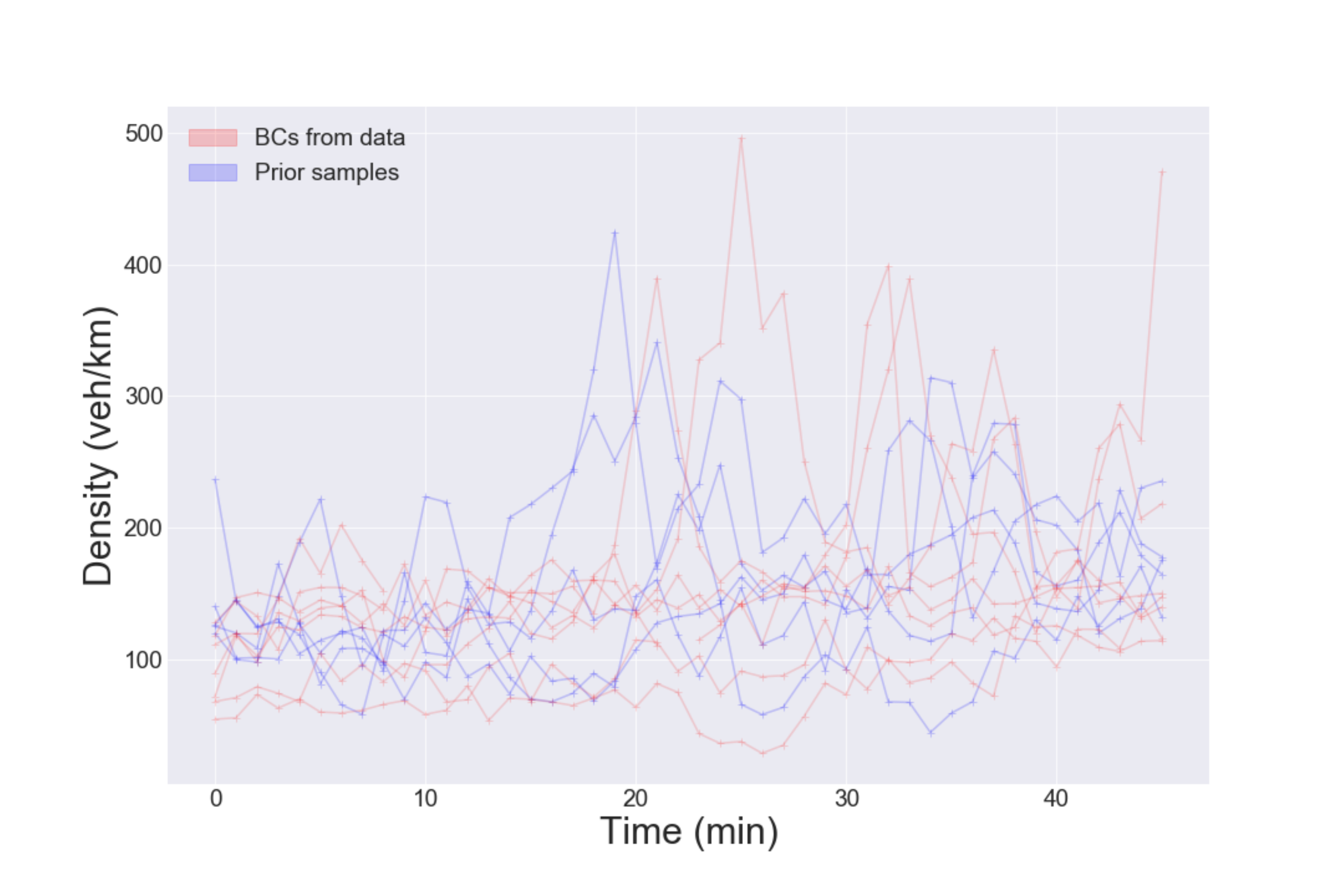}
  \caption{Inlet BC prior samples vs data}
  \label{BConly-prior_vs_data}
\end{subfigure}%
\begin{subfigure}{.5\textwidth}
  \centering
\includegraphics[width=1\linewidth]{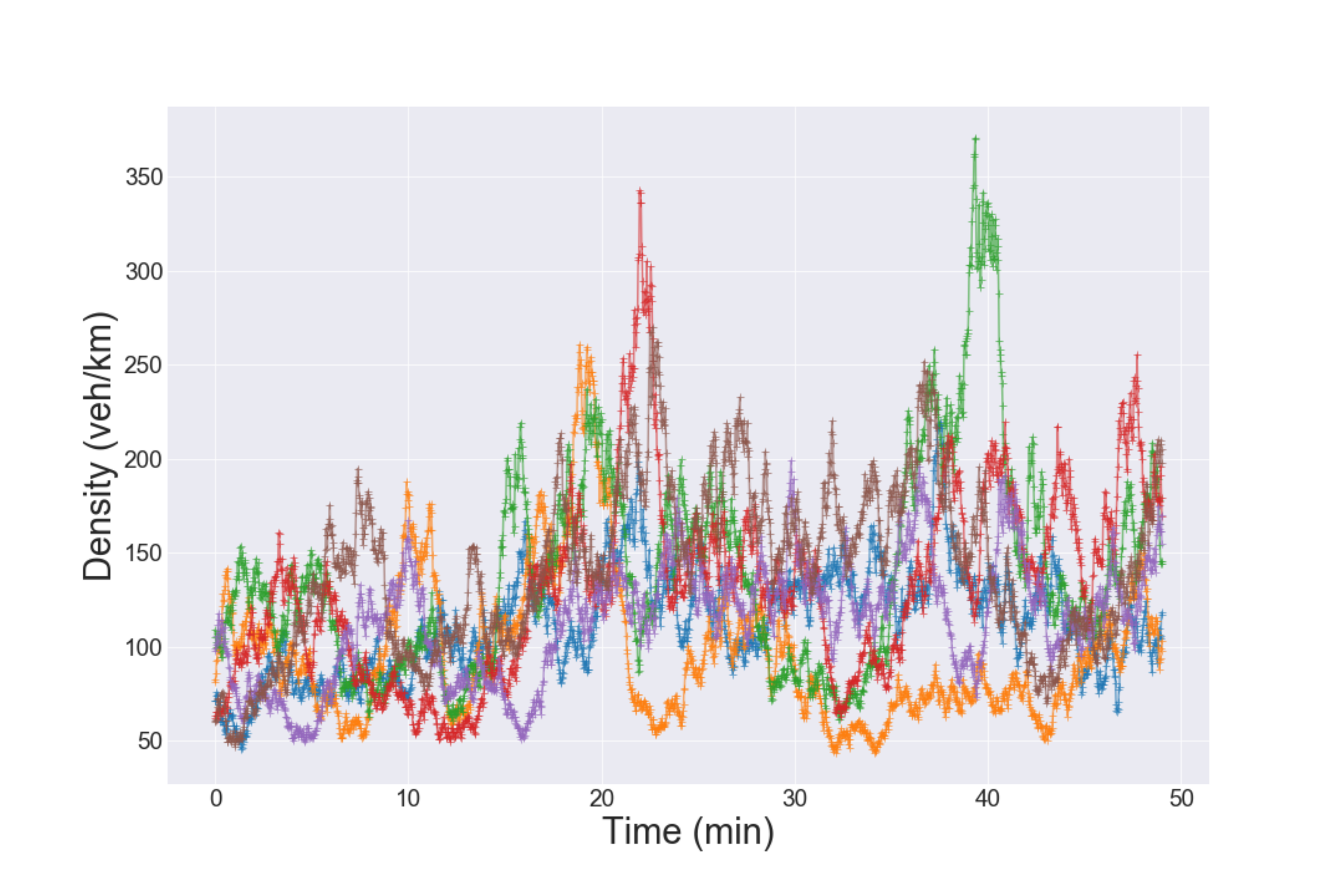}
  \caption{Inlet BC prior samples - high resolution}
  \label{BConly-prior_high_res}
\end{subfigure}
\captionsetup{width=.7\linewidth}
\caption{(a) Inlet BCs from data (using density from occupancy) along with prior samples at 1 min resolution (b) Samples from the prior for the inlet at full resolution: 1 point every 1.5 seconds}
\label{BConly-prior_vs_data_high_res}
\end{figure}

We plot in figure \ref{BConly-prior_vs_data} inlet BCs from data (data used to fit the OU process) along with prior samples. To allow comparison to the BCs from data, the prior samples here have the same resolution: one point per minute. We can visually check that the log-OU prior fits fairly well the inlet BCs from data (prior samples for the outlet BCs are also similar to BCs from data).

We also plot in figure \ref{BConly-prior_high_res} samples from the inlet BCs at full resolution: one point every 1.5 seconds, which is the resolution that we use in the inference. We use this resolution as it allows for a detailed description of the density waves that will get propagated by LWR. Indeed comparing the two figures, we can see that a resolution of 1min does not capture the details of the high density waves.

\subsection{Treatment of the initial conditions}

LWR requires the initial condition as well as the boundary conditions. The density will be propagated with either free flow or congested flow wave speed and so only the density measured for the first few minutes will be influenced by the initial condition. To avoid having to infer this initial condition, we simply do not use these first few detector times to build the likelihood; as a result the likelihood is unaffected by the initial condition and is only affected by the choice of boundary conditions and FD parameters.
To be able to only remove a small number of points in the \(x-t\) plane (ie: just the first few minutes) we assume that the FD parameters are such that density for these initial times corresponds to free flow (which is a reasonable assumption as can be seen in figure \ref{intro_M25_data}). We further assume that the free flow wave speed lies within a reasonable range of speeds. We remove the influence of the initial condition on the likelihood in this way for all further inferences in the paper.


\section{Results and discussion}
\label{sec:results}

\subsection{Sampling results}

We run the sampler for $102,000$ iterations and thin the samples by $100$.

\begin{figure}[ht]
  \centering
\includegraphics[width=.6\linewidth]{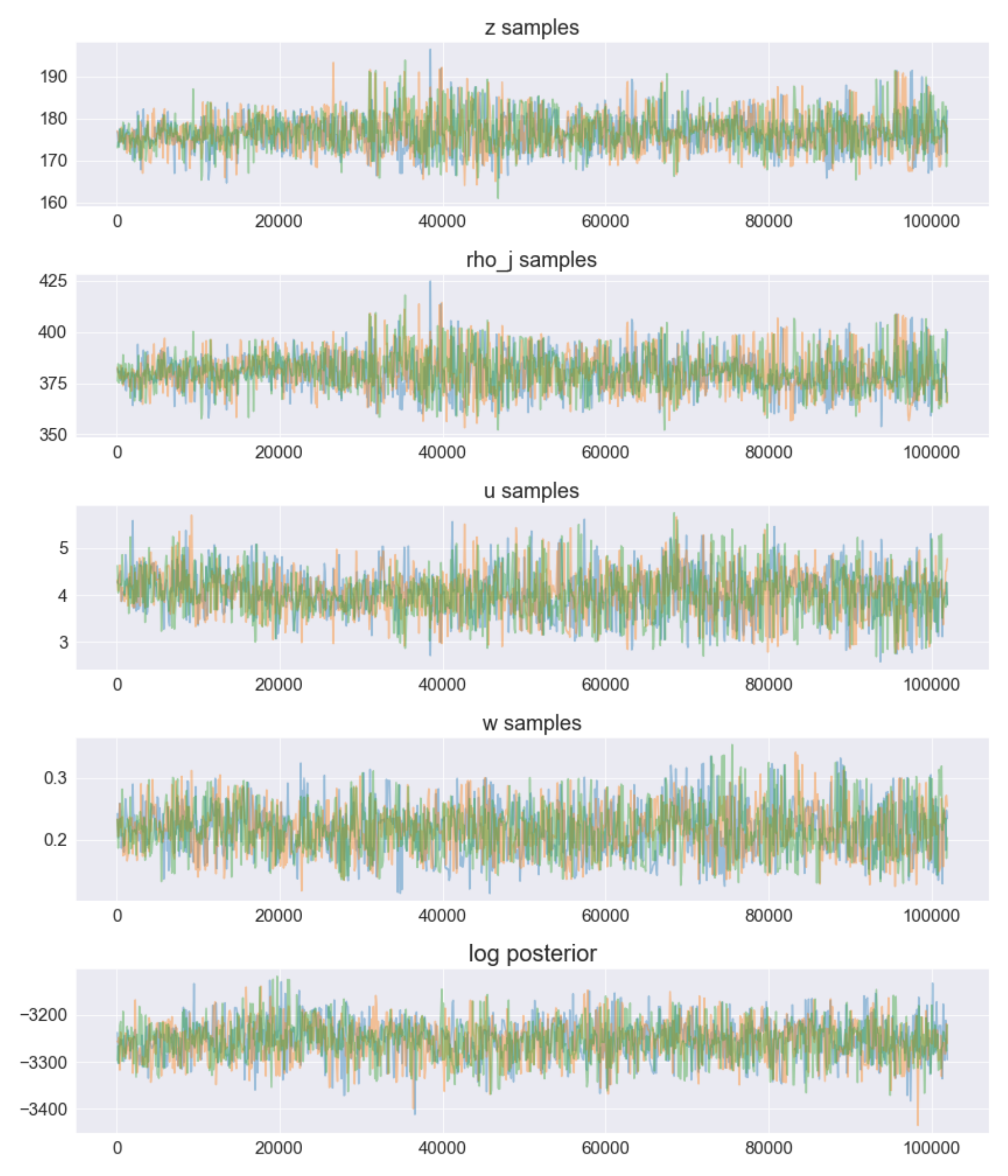}
  \caption{Trace plots for the FD parameters for a PT FES sampler, which show good mixing.} 
  \label{FDBC-FD_trace_plots}
\end{figure}

\begin{figure}[ht]
  \centering
\includegraphics[width=.7\linewidth]{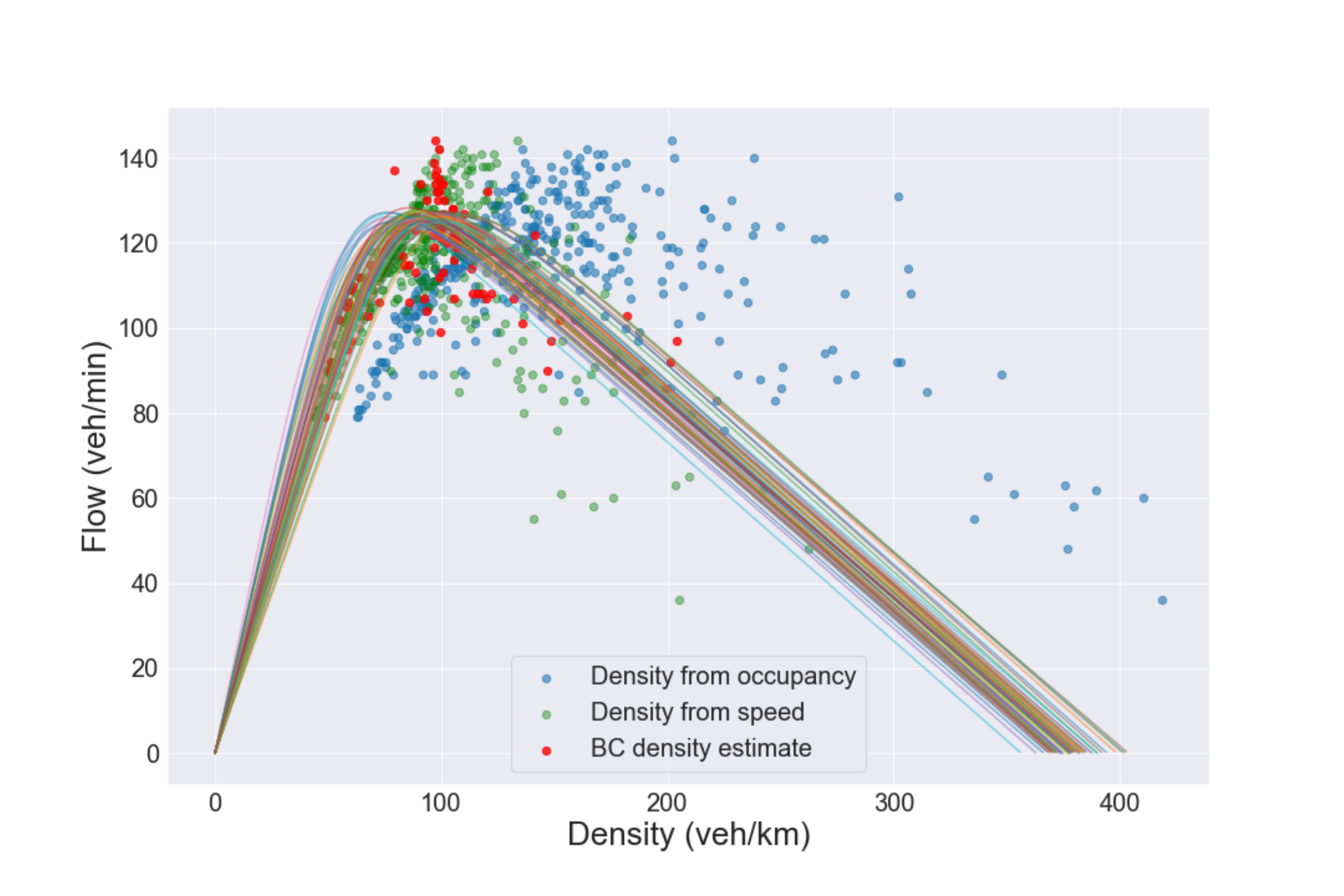}
  \caption{FD samples plotted with M25 flow data and three density estimation methods: from occupancy, from speed, and from BCs. The samples are from FD and BC sampling for del Castillo's FD for M25 data. The density estimated in the BCs seems to agree with density from speed, but the congested flow wave speed in the fitted model seems to be different to the wave speeds implied by the other two density estimation methods} 
  \label{FDBC-FD_samples}
\end{figure}

We show in figure \ref{FDBC-FD_trace_plots} the trace plots for the FD parameters and for a few of the walkers which show good mixing. We show in figure \ref{FDBC-FD_samples} the FD samples plotted with M25 flow data against 3 density estimates: density from speed, from occupancy, and estimated in the BCs. To obtain the latter we used the mean BCs (both inlet and outlet) and picked out the time points that correspond to measurements. We then plotted the M25 flow data at those time points against the density in the BC means. We first observe that for a given value of flow, the densities estimated in the BCs do not agree with densities estimated from occupancy but somewhat agree with the densities estimated from speed. In terms of the wave speeds, the free flow wave speeds implied by all three density estimates seem to agree, but the congested wave speeds do not. The congested flow wave speed in the fitted model seems to be in between the wave speed implied by the two other density estimates. 

\begin{figure}
\centering
\begin{subfigure}{.5\textwidth}
  \centering
\includegraphics[width=1\linewidth]{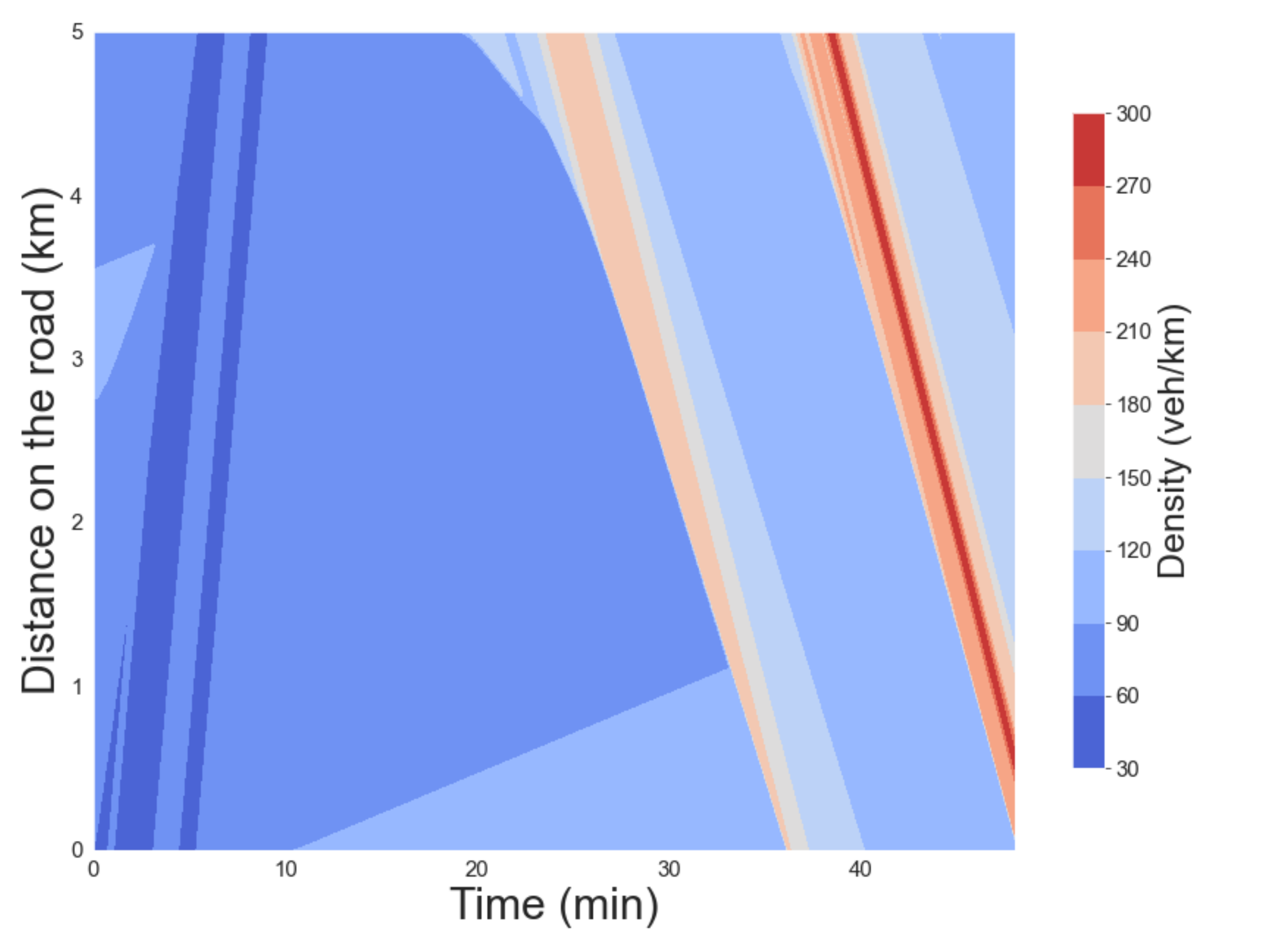}
  \caption{\(x-t\) plane from posterior mean}
  \label{FDBC-XT}
\end{subfigure}
\begin{subfigure}{.5\textwidth}
  \centering
\includegraphics[width=1\linewidth]{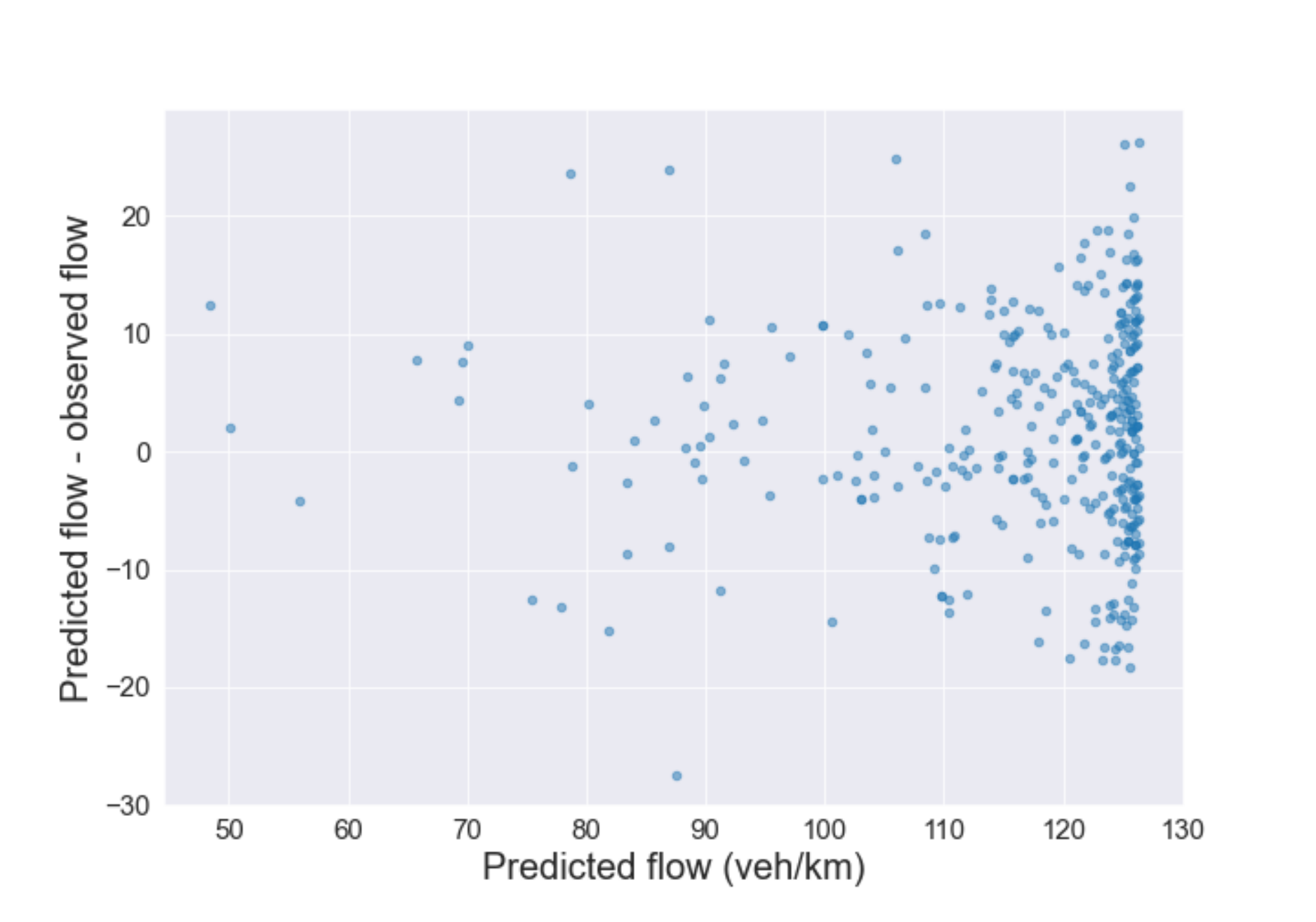}
  \caption{Residuals from posterior mean}
  \label{FDBC-residuals}
\end{subfigure}
\captionsetup{width=.7\linewidth}
\caption{Using the posterior mean parameters from FD and BC sampling with a PT FES sampler, we plot the output of LWR in the \(x-t\) plan in figure $a$ and the residuals in figure $b$}
\label{FDBC-XT_and_residuals}
\end{figure}

As the congested flow waves in the fitted model (in figure \ref{FDBC-XT}) seem to agree with the waves in M25 data (in figure \ref{intro_M25_data}) , this suggests that estimating the density in LWR rather than estimating it in a preprocessing step yields a better fit of the wave speeds. Finally, we show the residuals in figure \ref{FDBC-residuals} which suggests a good model fit.

\subsection{Discussion on the correlations and multi-modality}
\label{sec:multimodality}

The slow mixing speed can in part be explained by the strong correlations between the outlet and inlet BC parameters: we can see in figure \ref{intro_M25_data} that a forwards moving free flow wave leaving the inlet BC (with wave speed given by the slope of the FD) will hit the outlet BC (and vice versa for congested flow waves). Changing one of the BCs therefore requires changing the other one in a way that is compatible with the first. Furthermore, as hyperbolic PDEs are prone to shock formation, the BCs will have sharp changes in density: a small translation of the BC to the left or right (say) will therefore cause a large drop in likelihood. In  contract, PDEs that exhibit diffusion will smooth out such discontinuities quickly, so such a misfit will be penalised much less by the likelihood. An algorithm such as pCN - which is simply random walk proposal - will therefore mix slowly.

\begin{figure}
\centering
\begin{subfigure}{.5\textwidth}
  \centering
\includegraphics[width=1\linewidth]{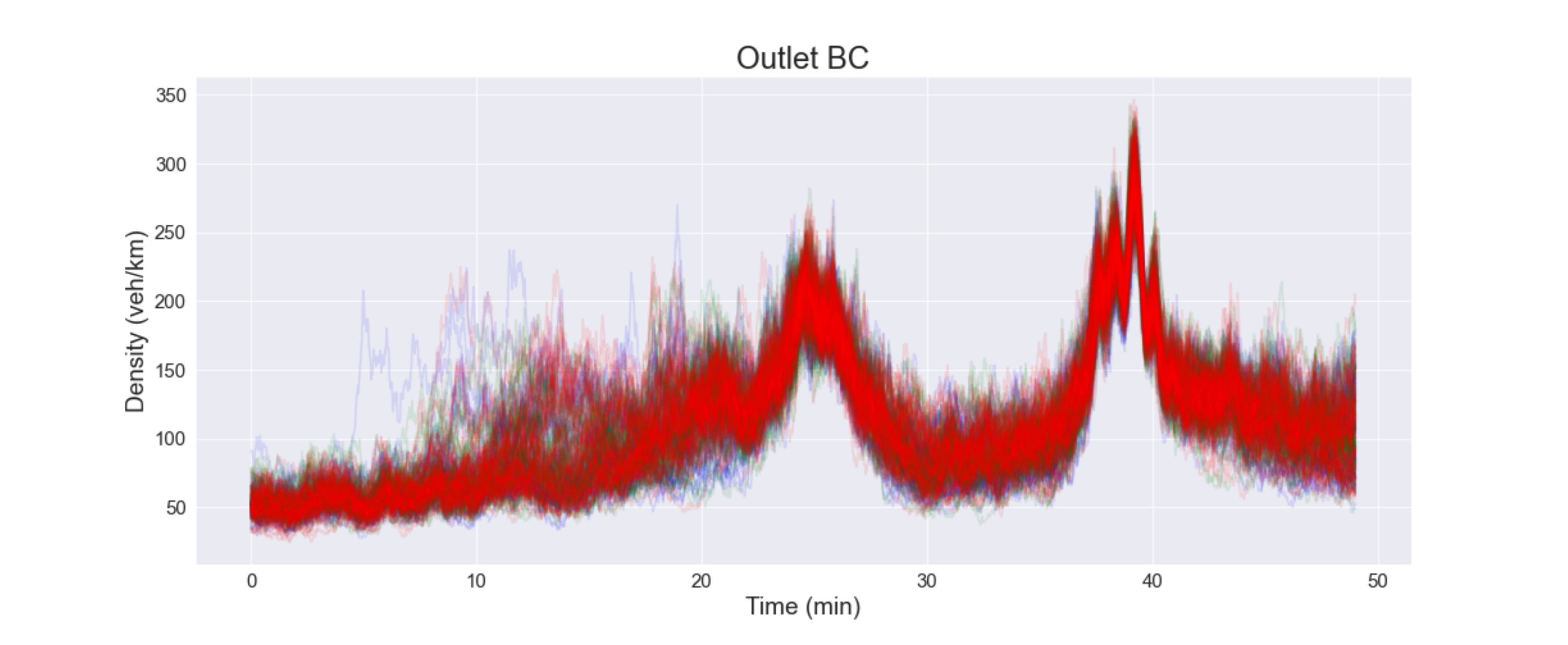}
  \caption{outlet BC trace plots}
  \label{FDBC-outlet_samples}
\end{subfigure}%
\begin{subfigure}{.5\textwidth}
  \centering
\includegraphics[width=1\linewidth]{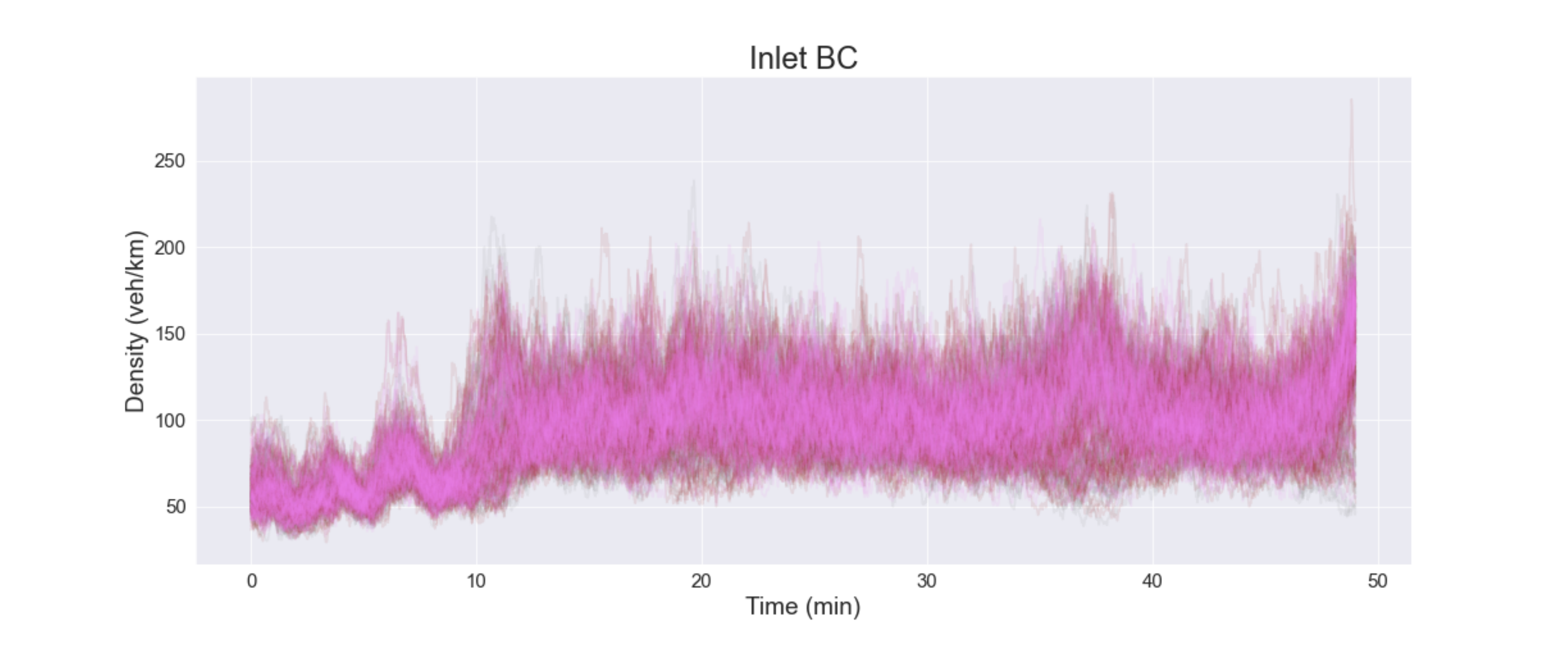}
  \caption{Inlet BC trace plots}
  \label{FDBC-inlet_samples}
\end{subfigure}
\captionsetup{width=.7\linewidth}
\caption{BC samples for the population PT sampler sampling FD and BC parameters.}
\label{FDBC-BC_samples}
\end{figure}

\begin{figure}[ht]
  \centering
\includegraphics[width=.7\linewidth]{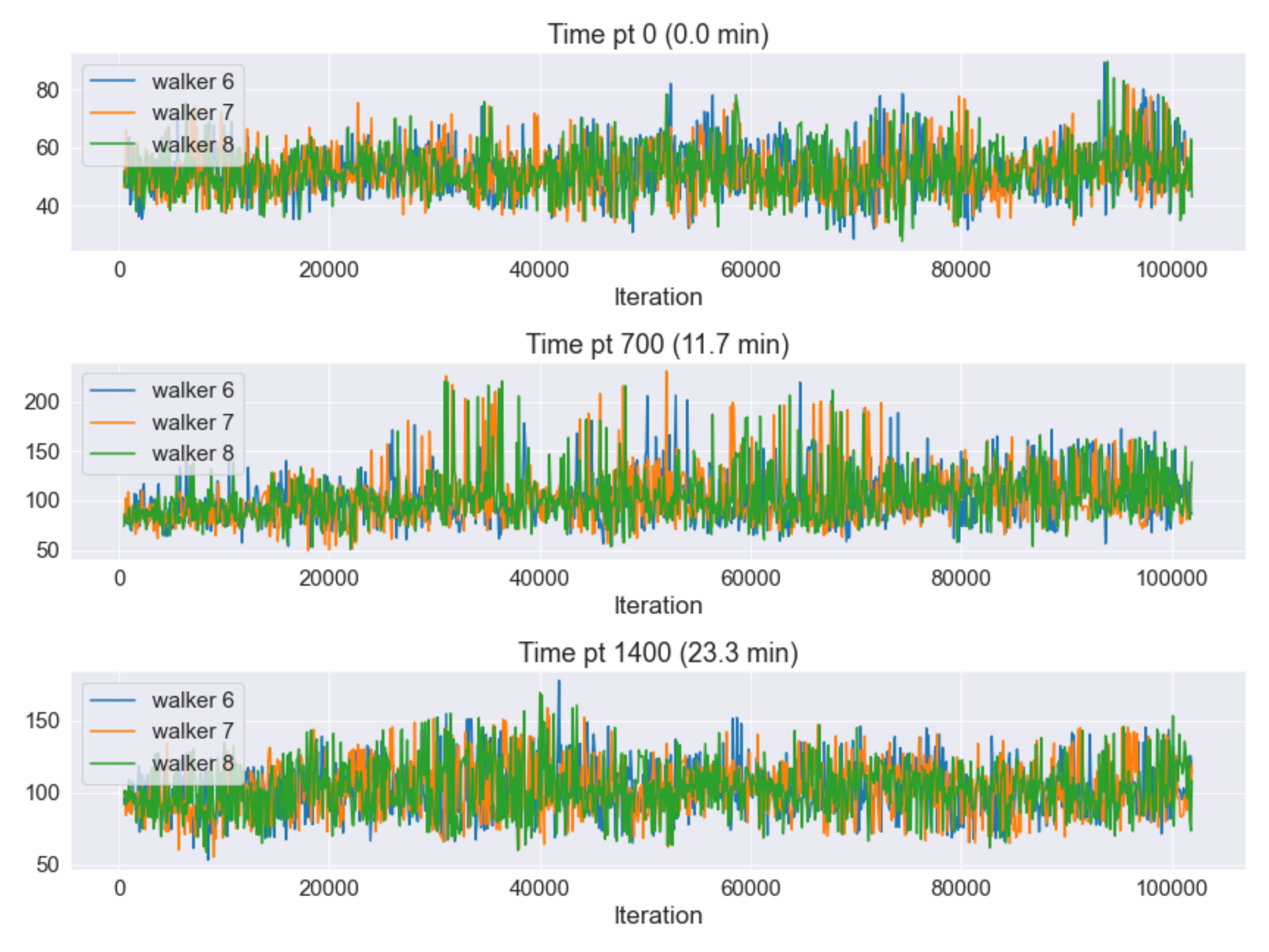}
  \captionsetup{width=.7\linewidth}
  \caption{Trace plots for 3 time points in the outlet BC which show some of the multimodality} 
  \label{FDBC-BC_outlet_trace_plots}
\end{figure}

\begin{figure}[ht]
  \centering
\includegraphics[width=.6\linewidth]{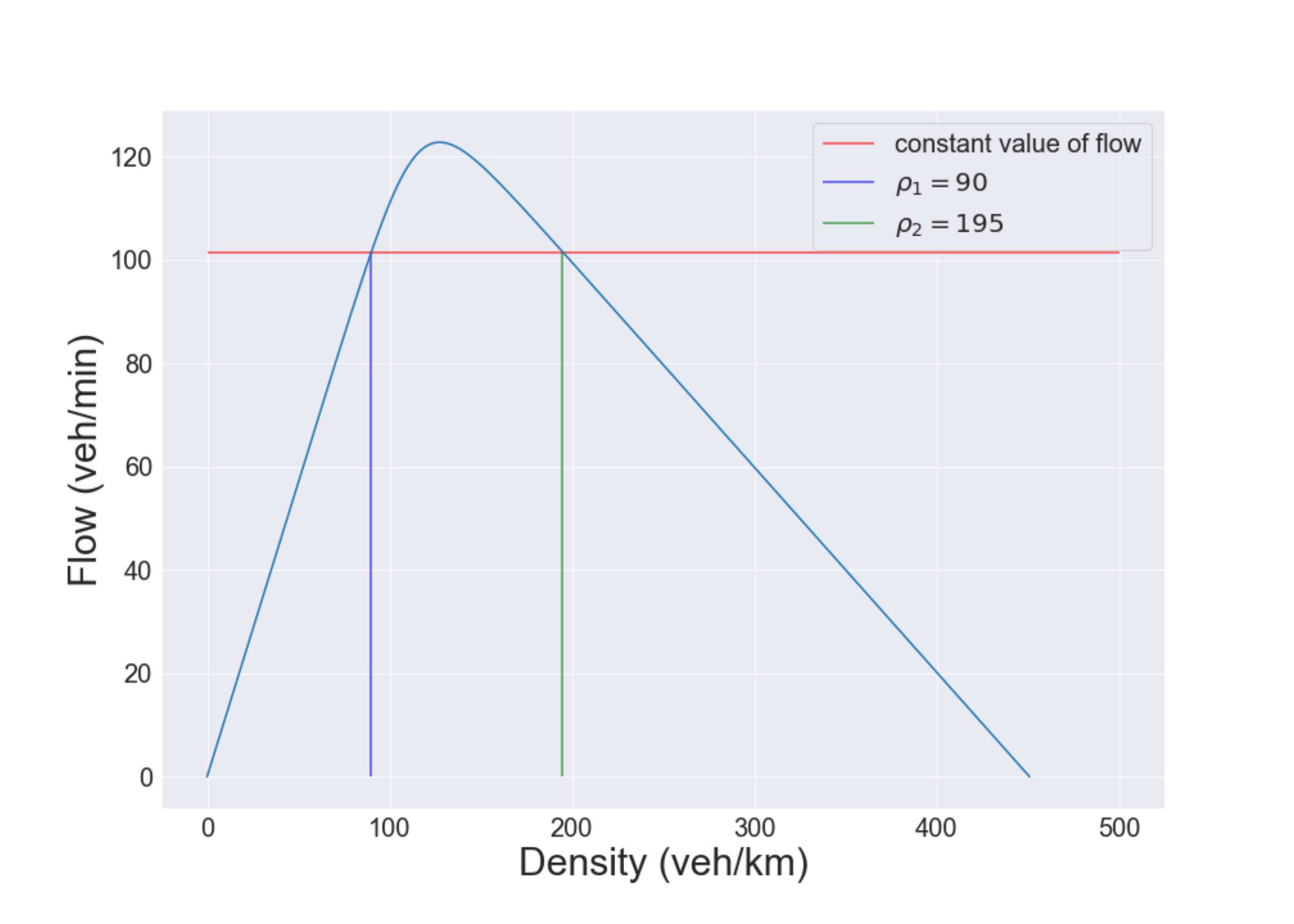}
  \captionsetup{width=.7\linewidth}
  \caption{Del Castillo FD. The two vertical lines correspond to two values of density (\(\rho_1=90\) and \(\rho_2=195\)) that map to the same value of flow. As the likelihood is built from flow, these two values of density are equally likely and therefore the posterior exhibits multi-modality.} 
  \label{BConly_keep_flow_constant}
\end{figure}

To explain the multi-modality visible in figures \ref{FDBC-outlet_samples} and \ref{FDBC-BC_outlet_trace_plots}, we recall that the likelihood is built from flow. This means that different values of density that map to the same value of flow will be equally likely. To illustrate this, we plot in figure \ref{BConly_keep_flow_constant} del Castillo's FD with the same parameters values used in the sampler, and we plot two vertical lines for two values of density (\(\rho_1=90\) and \(\rho_2=195\)) that map to the same value of flow (the horizontal line). If we then inspect the sections of the outlet and inlet BCs that exhibit multi-modality, we observe that some of the pairs of density branches approximately correspond to these two values. Of course there is dynamic behaviour in the \((x-t)\) plane so this explanation is an approximation.

\section{Conclusion}

After having given a brief review of motorway traffic flow modelling, we fit the Fundamental Diagram (FD) directly to flow-density data, but found that estimating the FD and BCs with LWR as the forward model resulted in a superior fit in terms of wave speed. The fitting procedure required a state-of-the-art gradient free sampler augmented with parallel tempering to sample from the highly correlated and multi-modal posterior.

We have provided a unified statistical model to estimate both boundary conditions and fundamental diagram parameters while respecting the character of LWR as a conservation law. Furthermore, we compared the density estimated in the BCs to two density estimates in the engineering literature (density from occupancy and speed) and find that although the free flow wave speeds implied by the three methods agree, only the congested flow wave speeds in the density from BCs (namely, in the fitted model) agree with the congested flow waves in M25 data. When inserted into LWR, the BCs estimated using our method provides a fit superior to that obtained from BCs using engineering methods.

Furthermore, the sampling methodology developed in this paper could be used to fit more sophisticated traffic flow models to data. Examples of these are systems of PDEs that include conservation of momentum as well as mass (namely, 2 equation models), which are discussed in \cite{Coullon_phd}. These classes of models could be quantitatively compared after fitting the parameters and boundary conditions for each of them. This would allow for a rigorous assessment of the strengths and weaknesses of these models.


\clearpage

Replication data and code can be found here: \href{https://github.com/jeremiecoullon/BIP\_LWR-paper}{https://github.com/jeremiecoullon/BIP\_LWR-paper}

\bibliographystyle{apalike}

\begin{thebibliography}{}

\bibitem[Atchade et~al.(2011)]{Atchade_optimal_scaling}
\textbf{Atchad\'e, YF, Roberts, GO, Rosenthal, JS} (2011) Towards optimal scaling of metropolis-coupled Markov chain Monte Carlo, \emph{Statistics and Computing} 21(4), pp. 555-568.

\bibitem[Aw and Rascle(2000)]{A&R_resurection_of_fluid}
\textbf{Aw, A, Rascle, M} (2000) Resurection of "second order" models of traffic flow. \emph{SIAM Journal on Applied Mathematics}, 60(3), pp.916-938.

\bibitem[Bando et~al.(1995)]{Bando1995}
\textbf{Bando, M, Hasebe, K, Nakayama, A, Shibata, A, and Sugiyama} (1995). Dynamic Model of traffic congestion and numerical simulation. \emph{Physical Review E} 51, 1035-1042. (doi:10.1103/PhysRevE.51.1035)

\bibitem[Beskos et~al.(2017)]{Beskos_geometric_MCMC_2017}
\textbf{Beskos, A, Girolami, M, Lan, S, Farrell, PE, Stuart, AM} (2017) Geometric MCMC for infinite-dimensional inverse problems \emph{Journal of Computational Physics} 335, pp. 327-351 doi: https://doi.org/10.1016/j.jcp.2016.12.041

\bibitem[Bonzani abd Gramani(2009)]{Bonzani_review}
\textbf{Bonzani, I, Gramani CLM} (2009) Critical analysis and perspectives on the hydrodynamic approach for the mathematical theory of vehicular traffic. \emph{Mathematical and Computer Modelling}, 50 526-541. doi:10.1016/j.mcm.2009.03.007

\bibitem[Brooks et~al.(2011)]{Handbook_of_MCMC}
\textbf{Brooks, S, Gelman, A, Jones, GL, Meng, X} (2011) \emph{Handbook of Markov Chain Monte Carlo}. CRC Press 

\bibitem[Calderhead(2014)]{Calderhead_parallelise}
\textbf{Calderhead B} (2014) A general construction for parallelizing Metropolis- Hastings algorithms. \emph{Proceedings of the National Academy of Sciences}, 111(49):17408- 17413, 2014.

\bibitem[Clawpack Development Team(2018)]{Clawpack_software}
\textbf{Clawpack Development Team} (2018) Clawpack Version 5.4.0, http://www.clawpack.org, doi:10.5281/zenodo.262111

\bibitem[Cotter et~al.(2009)]{BIP_fluids}
\textbf{Cotter, SL, Dashti, M, Robinson, JC, Stuart, AM} (2009) Bayesian inverse problems for functions and applications to fluid mechanics. \emph{Inverse Problems} \textbf{25}. doi:10.1088/0266-5611/25/11/115008

\bibitem[Cotter et~al.(2013)]{Cotter_pCN}
\textbf{Cotter, SL, Roberts GO, Stuart, AM, White, D} (2013) MCMC Methods for Functions: Modifying Old Algorithms to Make Them Faster. \emph{Statistical Science} 28(3), pp. 424-446 DOI: 10.1214/13-STS421

\bibitem[Coullon(2019)]{Coullon_phd}
\textbf{Coullon, J} (2019) MCMC for a hyperbolic Bayesian inverse problem in motorway traffic flow. \emph{PhD dissertation}. University College London.

\bibitem[Coullon and Webber (2021)]{Coullon_FES}
\textbf{Coullon, J and Webber RJ} (2021) Ensemble sampler for infinite-dimensional inverse problems. \emph{Statistics and Computing} 31(28) doi: https://doi.org/10.1007/s11222-021-10004-y


\bibitem[Cui et~al.(2014)]{Cui_LIS}
\textbf{Cui, T, Martin, J, Marzouk, YM, Solonen, A, and Spantini, A} (2014) Likelihood-informed dimension reduction for nonlinear inverse problems. \emph{Inverse Problems} 30(11)


\bibitem[Daganzo(1994)]{Daganzo_CT_model}
\textbf{Daganzo, CF} (1994) The cell transmission model: a dynamic representation of highway traffic consistent with the hydrodynamic theory. \emph{Transportation Research Part B: Methodological}, 28(4) pp. 269-287.

\bibitem[del Castillo(2012)]{del_Castillo_FD_paper}
\textbf{del Castillo, J} (2012) Three new models for the flow-density relationship: derivation and testing for freeway and urban data. Transportmetrica 8(6): 443-465

\bibitem[Gasser et~al.(2004)]{Gasser2004}
\textbf{Gasser, I, Sirito, G, and Werner, B} (2004) Bifurcation analysis of a class of “car following” traffic models. \emph{Physica D} 197, p.222-241.


\bibitem[Goodman and Weare(2010)]{Goodman_2010}
\textbf{Goodman J and Weare J} (2010) Ensemble samplers with affine invariance \emph{Communications in Applied Mathematics and Computational Scienc} 5(1), p.65-80.



\bibitem[Graham and Storkey(2017)]{continuously_tempered_HMC}
\textbf{Graham, MM, Storkey, AJ} (2017) Continuously tempered Hamiltonian Monte Carlo. \emph{arxiv} Available at https://arxiv.org/abs/1704.03338v1 [Accessed 5th March 2019]


\bibitem[Greenshields(1934)]{Greenshields_1934}
\textbf{Greenshields BD} (1934) The photographic method of studying traffic behaviour. \emph{Proceedings of the 13th Annual Meeting of the Highway Research Board} pp. 382-399.

\bibitem[Heydecker and Addison(2011)]{Heydecker_estimate_AVL}
\textbf{Heydecker, BG, Addison, JD} (2011) Measuring traffic flow using real-time data.  In: Kuehne, R and Gartner, NA, (eds.) \emph{Transportation Research Circular, E-C149: 75 Years of the Fundamental Diagram for Traffic Flow Theory}. (pp. 109-120). 


\bibitem[Hoogendoorn and Bovy(2001)]{Hoogendoorn2001}
\textbf{Hoogendoorn, SP and Bovy, PHL} (2001) State-of-the-art of vehicular traffic flow modelling. \emph{Proceedings of the Institution of Mechanical Engineers, Part 1: Journal of Systems and Control Engineering} 215:283. doi: 10.1177/095965180121500402


\bibitem[Iacus and Stefano(2008)]{simulation_inference_SDE}
\textbf{Iacus, M, Stefano} (2008) \emph{Simulation and Inference for Stochastic Differential Equations}. Springer.


\bibitem[Iglesia  et~al.(2014)]{BIP_Darcy_flow}
\textbf{Iglesia, MA, Lin, K, Stuart, AM} (2014) Well-posed Bayesian geometric inverse problems arising in subsurface flow. \emph{Inverse Problems} \textbf{30}. doi:10.1088/0266-5611/30/11/114001


\bibitem[Jasra et~al.(2007)]{Population_based_MCMC_review}
\textbf{Jasra, A, Stephens, DA, Homes, CC} (2007) On population-based simulation for static inference. \emph{Statistics and Computing} 17(3), pp. 263-279.

\bibitem[Law(2014)]{operator_weighted_proposal_paper}
\textbf{Law, KJH} (2014) Proposals which speed up function-space MCMC. \emph{Journal of Computational and Applied Mathematics} \textbf{262}(15) pp 127-138


\bibitem[Leveque(2004)]{FVMHP_book}
\textbf{LeVeque RJ} (2004) \emph{Finite Volume Methods for Hyperbolic Problems}. Cambridge University Press.


\bibitem[Lighthill and Richards(1955)]{Lighthill_whitham}
\textbf{Lighthill, MH, and Whitham, GB} (1955) On Kinematic Waves 2: A Theory of Traffic Flow on Long Crowded Roads. Proceedings of the Royal Society of London A, 229, pp.317-345.

\bibitem[Liu(2001)]{Liu_MC_strategies}
\textbf{Liu, JS} (2001) \emph{Monte Carlo Strategies in Scientific Computing}. Springer-Verlag New York

\bibitem[Mahnke et~al.(2005)]{stochastic_process_review}
\textbf{Mahnke, R, Kaupuzs, J, Lubashevsky, I} (2005) Probabilistic description of traffic flow. \emph{Physics Reports}, 408(2005), 1-130.


\bibitem[Mandli et~al.(2016)]{Clawpack_paper_citation}
\textbf{Mandli, KT, Ahmadia, AJ, Berger, MJ, Calhoun, DA, George, DL,
Hadjimichael, Y, Ketcheson, DI, Lemoine, GI, LeVeque, RJ} (2016)
Clawpack: building an open source ecosystem for solving hyperbolic PDEs.
PeerJ Computer Science. doi:10.7717/peerj-cs.68


\bibitem[Martino et~al.(2016)]{O-MCMC}
\textbf{Martino, L, Elvira, V, Luengo, D, Corander, J, Louzanda, F} (2016) Orthogonal parallel MCMC methods for sampling and optimization. \emph{Digital Signal Processing}, 58 (2016), 64-84.

\bibitem[Polson and Sokolov(2015)]{Polson_LWR}
\textbf{Polson, N, Sokolov, V} (2015) Bayesian analysis of traffic flow on interstate I-55: the LWR model. \emph{The Annals of Applied Statistics} \textbf{9}(4), pp 1864-1888 


\bibitem[Richards(1956)]{Richards}
\textbf{Richards, PI} (1956) Shock Waves on the Highway. \emph{Operations Research}, 4 (1) pp.42-51

\bibitem[Sambridge(2014)]{Geophysics_PT}
\textbf{Sambridge, M} (2014) A Parallel Tempering algorithm for probabilistic sampling and multimodal optimization. \emph{Geophysical Journal International} \textbf{196} pp 357-374. DOI: doi: 10.1093/gji/ggt342


\bibitem[Stuart(2010)]{Stuart_Acta_Numerica}
\textbf{Stuart, AM} (2010) Inverse problems: A Bayesian perspective. \emph{Acta Numerica}, 19, pp 451-559 doi:10.1017/ S0962492910000061

\bibitem[Sokal(1989)]{Sokal_lecture_notes}
\textbf{Sokal, AD} (1989) \emph{Monte Carlo Methods in Statistical Mechanics: Foundations and New Algorithms}, lecture notes (unpublished). In \emph{Cours de Troisieme Cyle de la Physique en Suisse Romande}., Lausanne.


\bibitem[Tawn and Roberts(2018)]{Quant_TA_paper}
\textbf{Tawn, NG, Roberts, GO} (2018) Accelerating Parallel Tempering: Quantile Tempering Algorithm (QuanTA). \emph{arxiv} [preprint] Stat.ME. Available at: https://arxiv.org/abs/1808.10415v1 [Accessed 5th March 2019]

\bibitem[Hohage and Werner(2016)]{Hohage_Werner_Poisson_inverse}
\textbf{Hohage T, Werner F} (2016) Inverse Problems with Poisson Data: statistical regularization theory, applications and algorithms. Inverse Problems 32: 093001:56pp.

\bibitem[Ward(2009)]{Ward_phd}
\textbf{Ward, J} (2009) \emph{Heterogeneity, Lane-Changing and Instability in Traffic: a Mathematical Approach}. PhD dissertation. University of Bristol.


\bibitem[Zhang(2000)]{Zhang2}
\textbf{Zhang, HM} (2000) A non-equilibrium traffic model devoid of gas-like behaviour. \emph{Transportation Research Part B}, 36 (2002) pp.275-290.








\end{thebibliography}

\begin{appendices}

\section{Direct fit: FD sampling}
\label{appendix-FD_sampling}

Covariance matrix for the direct fit to data for \(\begin{pmatrix}z & \rho_j  & u & w\end{pmatrix}^T\) :

\(\begin{pmatrix}
182.292318 & -288.07905 &  -2.34389543 &  1.21897887 \\
-288.07905 & 561.808314 &  5.26749447 &  -1.7470824 \\
-2.34389543 &    5.26749447 & 0.08204741 &  -0.00839764 \\
1.21897887 & -1.7470824 & -0.00839764 &  0.00931977
\end{pmatrix}\)

\section{FES with PT}
\label{appendix-FES_PT}

\begin{itemize}
    \item Metropolis-within-Gibbs move probabilities for AIES, pCN oulet, pCN inlet swap: $[0.25, 0.125, 0.125, 0.5]$
    \item inverse-Temperatures: $[1, 0.76, 0.58, 0.44]$
    \item pCN step sizes for outlet (for each inverse-temperature): $[0.078, 0.09, 0.11, 0.15]$ 
    \item pCN step sizes for inlet (for each inverse-temperature): $[0.155, 0.17, 0.2, 0.25]$
    \item FES truncation: $M_{trunc}=4$
\end{itemize}

\end{appendices}

\end{document}